\documentclass[%
 aip,
rsi,
floatfix,
 amsmath,amssymb,
 reprint,%
]{revtex4-2}

\usepackage[caption=false,font=normalsize,labelfont=sf,textfont=sf]{subfig}
\usepackage{graphicx}
\usepackage{hyperref}
\usepackage{url}
\usepackage{textgreek}

\newcommand{\rme}{\mathrm{e}}
\newcommand{\NISTdisc}{\footnote{\label{NISTdisclaimer}Certain commercial equipment, instruments, or materials (or suppliers, or software, ...) are identified in this paper to foster understanding. Such identification does not imply recommendation or endorsement by the \uppercase{N}ational \uppercase{I}nstitute of \uppercase{S}tandards and \uppercase{T}echnology, nor does it imply that the materials or equipment identified are necessarily the best available for the purpose.}}
\draft 

\begin{document}


\title{Quadrature amplitude modulation for electronic sideband Pound-Drever-Hall laser frequency locking} 



\author{J. Tu}
\author{A. Restelli}
\author{K. Weber}
\author{I. B. Spielman}
\author{S. L. Rolston}
\author{J. V. Porto}
\email{porto@umd.edu}
\author{S. Subhankar}
\email{sarthaks@terpmail.umd.edu}
\affiliation{Joint Quantum Institute, National Institute of Standards and Technology and the University of Maryland, College Park, Maryland 20742 USA}


\date{\today}

\begin{abstract}
The Pound--Drever--Hall (PDH) technique is routinely used to stabilize the frequency of a laser to a reference cavity. Electronic sideband (ESB) locking, a PDH variant, bridges the frequency gap between the discrete cavity resonances and a desired laser frequency. Here we use quadrature amplitude modulation (QAM), a standard technique in digital communications, to generate the high-quality phase-modulated radio-frequency (rf) drive required for ESB locking. We develop a theoretical framework to analyze how in-phase/quadrature-phase (I/Q) impairments distort the ESB error signal and induce frequency offsets relevant to ultranarrow-linewidth lasers. We then design and implement a direct software-defined radio (SDR) on an UltraScale+ RFSoC platform, frequently adopted across modern quantum-computing systems, to digitally compensate QAM I/Q impairments. Using this device, we generate phase-modulated rf signals with a large phase-modulation index of $1.01$~rad and root-mean-square I/Q errors below $0.3\ \%$ over a carrier-frequency range of $350~\mathrm{MHz}$ to $1.75~\mathrm{GHz}$. Finally, we lock a laser to an ultralow expansion (ULE) reference cavity and demonstrate continuous laser-frequency tuning by ramping the carrier frequency while maintaining lock, validating the continuous tunability of our ESB locking instrument.
\end{abstract}


\maketitle 

\section{\label{sec:Introduction}Introduction}
Ultranarrow-linewidth lasers are used in many areas of physics and engineering. Examples include precision spectroscopy~\cite{PhysRevResearch.1.033113,PhysRevA.107.L060801}, optical atomic clocks~\cite{Ludlow2015,Katori2011,Derevianko2011,PhysRevA.98.022501,PhysRevLett.95.083003}, quantum computation and simulation~\cite{Levine2018,DeLeseleuc2018,Graham2019}, gravitational wave detection~\cite{Izumi2014,Kapasi2020,Buikema2019}, fiber optic sensing~\cite{Elsherif2022,Bao2012}, and light detection and ranging (i.e., LIDAR)~\cite{Bai2021,Tang2022,Carlson2009,Sun2019}. Narrow-linewidth operation is routinely achieved by actively stabilizing the frequency of a free-running laser to a high-finesse optical cavity using the Pound--Drever--Hall (PDH) locking scheme~\cite{Drever1983,Black2001}. In the standard PDH scheme, the laser is locked to a cavity resonance, so the available lock points are separated by integer multiples of the cavity free spectral range (FSR). As a result, an additional frequency offset must be provided whenever the desired laser frequency lies between adjacent cavity modes.

Acousto-optic modulators (AOMs) can provide frequency offsets, but their tuning range is generally limited to a few hundred MHz—a small fraction of the GHz-scale FSR typical of ultralow expansion (ULE) reference cavities~\cite{Alnis2008,Zeng:21,Baryshev_2012}. Serrodyne frequency shifting using electro-optic modulators (EOMs) extends this tuning range from a few hundred MHz to over a GHz, but requires wide-bandwidth EOMs and high-precision control of the sawtooth waveform when the detuning is large~\cite{Houtz:09,Johnson:10,Kohlhaas:12,Hildebrand:25,Hildebrand:25-1}. Offset-locking a second laser to a reference laser enables tuning over hundreds of GHz, but at the significant cost of an additional laser system~\cite{Ye:99,Zhou:23}. Alternatively, the radio-frequency (rf) waveform driving the EOM used for PDH locking can be engineered to supply a tunable frequency offset. In this approach, known as offset-sideband locking, the cavity stabilizes a phase-modulation sideband that is offset from the optical carrier by a controllable rf frequency, rather than stabilizing the carrier itself. By varying this carrier--sideband frequency offset via the applied rf drive waveform, the carrier inherits the cavity stability while being tuned continuously over an rf bandwidth of a few GHz~\cite{Thorpe:08,Bai_2017,Milani:17,PhysRevLett.121.053001,Sanjuan:21}.

Two popular implementations of offset-sideband locking are electronic sideband (ESB) locking~\cite{Bai_2017,Milani:17} and dual-sideband (DSB) locking~\cite{PhysRevLett.121.053001,Sanjuan:21}. In the ESB scheme, the rf waveform applied to the EOM is $V_{\mathrm{EOM}} \propto \beta_c \sin\big(\Omega_c t + \beta_m \sin(\Omega_m t)\big)$, whereas in the DSB scheme the rf waveform is $V_{\mathrm{EOM}} \propto \beta_c \sin(\Omega_c t) + \beta_m \sin(\Omega_m t)$. Here $\Omega_c$ is the carrier--sideband frequency offset used to tune the stabilized laser, $\Omega_m$ is the fixed PDH demodulation frequency, and $\beta_c$ and $\beta_m$ are the corresponding modulation indices. In both schemes, the EOM generates an optical sideband centered at $\Omega_0\pm\Omega_c$ (with $\Omega_0$ the laser carrier) that is locked to the cavity resonance. The PDH error signal is obtained by demodulating at $\Omega_m$ using the sidebands at $\Omega_0+\Omega_c,\Omega_0+\Omega_c\pm\Omega_m.$ 
Varying $\Omega_c$ tunes the stabilized laser frequency continuously while maintaining the cavity-referenced linewidth. The DSB waveform is straightforward to generate, but it produces spurious sidebands that can introduce unintended lock points. By contrast, the ESB waveform is more demanding to synthesize, but it strongly suppresses spurious features, simplifying lock acquisition and doubling the tuning range relative to DSB~\cite{Livas_2009}. 

ESB locking has been used for two-photon excitation of atoms to Rydberg states~\cite{Bridge:16,Legaie:18} and precision laser spectroscopy~\cite{Rabga:23,Guttridge2018}. In these implementations, the required rf drive has typically been generated using analog rf mixers or analog in-phase/quadrature-phase (I/Q) modulator chips. Quadrature amplitude modulation (QAM), widely used in digital communications, provides a convenient framework for generating phase-modulated rf signals by independently controlling the in-phase and quadrature components with two baseband waveforms~\cite{Hanzo2011,Alencar2018}. For example, Ref.~\cite{Rabga:23} used QAM to implement an Armstrong-type phase-modulation approach to engineer the ESB rf signal~\cite{Armstrong1936}. This approach is well suited for low modulation index, whereas significant phase distortions can emerge at large modulation index. In ESB locking, modulation indices on the order of unity (e.g., $\beta_m\sim 1$ rad)~\cite{Thorpe:08} are often desirable to maximize the slope of the PDH error signal and thereby increase the achievable open-loop gain of the servo. In general, higher open-loop gain enables stronger suppression of frequency noise and thus narrower stabilized linewidth~\cite{Wang2025}.

A practical limitation of prior ESB implementations is the lack of a detailed treatment of how imperfections in the generated phase-modulated waveform affect the ESB error signal. Such imperfections can introduce unwanted frequency offsets in the ESB error signal, and these offsets may drift in time. In Ref.~\cite{Rabga:23}, the effects of these imperfections are inconsequential (and therefore likely not emphasized) for spectroscopy of a moderately narrow transition with linewidth $\Gamma/2\pi \approx 380~\mathrm{kHz}$, since small parasitic frequency offsets remain well within the transition linewidth. However, similar offsets and drift can become a significant error source in precision-metrology applications, such as probing ultranarrow optical transitions~\cite{Rosenband2007,Porsev2004,Muniz2021,Dolde2025,Nicholson2015,Takamoto2003,Derevianko2011}.

Motivated by the use of ESB locking in precision metrology, we develop a theoretical treatment of how errors in QAM-based ESB signal generation propagate into distortions and offsets in the ESB error signal. Unlike prior analog approaches, we synthesize the required phase-modulated rf waveform in the digital domain using an UltraScale+ RFSoC platform\cite{rfdc2024} that has been adopted in quantum-computing systems, including neutral-atom arrays~\cite{Evered2025,Maetani_2024}, superconducting circuits~\cite{Carobene_2025,10821231,PRXQuantum.5.020326,PRXQuantum.5.030347,10.1063/5.0076249}, and trapped ions~\cite{10845073}. Specifically, we generate the phase-modulated ESB waveform using QAM in a direct software-defined radio (SDR) configuration~\cite{booksdr} and convert it to an analog output using the on-chip, hardened RF digital-to-analog converter (RF-DAC), enabling portable single-board generation of high-quality phase-modulated rf signals for ESB.

We implement this signal-generation scheme on the RFSoC 4X2 kit (Real Digital and AMD)\NISTdisc and use it to generate ESB rf waveforms via QAM with controllable modulation imperfections (I/Q impairments).\footnote{GitHub repositories for the device can be found at \url{https://github.com/JQIamo/ESBL_RFSoc}.} Because the platform can both minimize intrinsic I/Q impairments and deliberately inject controlled impairments, it provides a versatile testbed for validating our analytic models and developing practical calibration routines. The digital approach reduces sensitivity to drift and retuning compared with analog signal chains, improving long-term stability.

The paper is organized as follows. Sec.~\ref{sec:theory} develops the theoretical model for QAM-based ESB signal generation and quantifies how I/Q impairments distort the ESB error signal and induce lock-point offsets. Sec.~\ref{sec:design} describes the RFSoC-based implementation of QAM waveform synthesis and impairment control. Sec.~\ref{sec:performance} experimentally verifies the fidelity of the ESB rf waveform through electronic measurements using a spectrum analyzer and optical tests in a PDH locking setup.

\section{\label{sec:theory}Theory}
Table~\ref{tab:my_label} summarizes the symbols used throughout this paper. This section develops the theoretical framework for ESB locking and QAM-based synthesis of the phase-modulated rf waveform. Sec.~\ref{sec:ESB-theory} introduces the ESB locking spectrum and shows how tuning the carrier frequency maps onto tuning the locked laser frequency. Sec.~\ref{sec:ESBDSB} compares ESB and DSB using the spectral-triplet picture, highlighting the suppression of spurious lock points and the doubled tuning range in ESB. Sec.~\ref{sec:QAM} presents QAM as a convenient method for synthesizing the ESB waveform. Sec.~\ref{sec:theory1} models I/Q impairments, connects them to measurable RMS I/Q magnitude and phase errors, and quantifies their impact on the ESB error-signal gain and offset. Finally, Sec.~\ref{sec:passband} discusses practical considerations that affect ESB performance, including bandwidth/dispersion limits and carrier-oscillator noise and drift.

\begin{table}[!h]
    \caption{\label{tab:my_label}List of symbols}
    \begin{ruledtabular}
    \begin{tabular}{ c  c } 
     Symbol & Description\\ \hline
     $E$ & phase-modulated optical field \\ 
     $\Omega_0$ & optical frequency of the laser  \\ 
     $\Omega_c$ & carrier rf frequency  \\ 
     $\Omega_m$ & baseband rf frequency  \\ 
     $\beta_c$ &  carrier modulation depth \\ 
     $\beta_m$ &  phase-modulation index \\ 
     $J_i$ &  Bessel function of the first kind for order $i$ \\ 
     $V_\textrm{EOM}$ & rf signal put into the EOM\\ 
     $I$ & in-phase baseband waveform \\ 
     $Q$ & quadrature baseband waveform \\ 
    $\xi$ & amplitude of $V_\textrm{EOM}$\\
     $\xi_I$, & in-phase baseband waveform amplitude\\
     $\xi_Q$, & quadrature baseband waveform amplitude\\
     $q$ & I/Q impairments such as $g$, $\phi$, $\Delta_I$, and $\Delta_Q$\\ 
     $g$ & gain imbalance of I/Q modulation \\ 
     $\phi$ & phase imbalance of I/Q modulation \\ 
     $\Delta_I$ & in-phase DC offset \\ 
     $\Delta_Q$ & quadrature-phase DC offset \\ 
     $\gamma$ & error signal gain \\ 
     $\delta$ & error signal offset \\ 
     $\kappa$ & optical cavity linewidth\\ 
     $s$ & instantaneous I/Q magnitude error\\
     $\zeta$ & instantaneous I/Q phase error\\
     $s_\textrm{RMS}$ & root-mean-square I/Q magnitude error\\
     $\zeta_\textrm{RMS}$ & root-mean-square I/Q phase error\\
    \end{tabular}
    \end{ruledtabular}
\end{table}

\subsection{ESB primer}\label{sec:ESB-theory}
The canonical optical electric field in ESB locking is
\begin{eqnarray}
    E(t)&=E_{\text{0}}\exp\left\{j\Omega_{0}t +j\underbrace{\beta_{c}\sin[\Omega_{c}t+\beta_{m}\sin(\Omega_{m}t)] }_{\propto V_{\textrm{EOM}}(t)}\right\}\label{eq:opticalelectricfield},
\end{eqnarray}
where $E_{\text{0}}$ is the amplitude of the optical electric field; $\Omega_{0}$ is the bare laser frequency; and $V_{\textrm{EOM}}(t)\propto\beta_{c}\sin[\Omega_{c}t+\beta_{m}\sin(\Omega_{m}t)]$ is the rf drive to the EOM phase-modulating the laser electric field. This rf signal $V_{\textrm{EOM}}(t)$ is parameterized by: \begin{itemize}
\item[$\beta_{c}$:] the carrier modulation depth;
\item[$\beta_{m}$:] the phase-modulation index;
\item[$\Omega_{c}$:] the carrier rf frequency [in the ultra high frequency (UHF) band];
\item[$\Omega_{m}$:] the baseband rf frequency [in the medium frequency (MF) or high frequency (HF) band].
\end{itemize} 
The carrier frequency $\Omega_{c}$ should be tunable approximately by the FSR of the cavity. 

The Fourier spectrum of the canonical optical electric field can be written, using the Jacobi--Anger identity, as
\begin{align}
E(t)
&= E_0 \sum_{n}\sum_{k} J_n\left(\beta_c\right)J_k\left(n\beta_m\right)
\mathrm{e}^{j\left(\Omega_0+n\Omega_c+k\Omega_m\right)t},
\label{deq:sidebandexpansion}
\end{align}
where $J_n(\cdot)$ denotes the Bessel function of the first kind of order $n$.
The amplitude of a sideband at frequency $\Omega_0 +n \Omega_c+k \Omega_m$ is $E_0J_n\left(\beta_c\right) J_k\left(n \beta_m\right)$.
In the ESB locking scheme~\cite{Thorpe:08}, the laser sideband at $\Omega_0\pm\Omega_{c}$ is locked to (and is therefore in resonance with) the cavity i.e. 
\begin{align}
    &\Omega_0\pm\Omega_c=2\pi N\times\textrm{FSR},
        \label{eq:offset}
\end{align}
where $N$ is the longitudinal mode index of the reference cavity and $\textrm{FSR}$ is the cavity free spectral range. 
Assuming that the laser stays locked to the chosen reference cavity mode, a $\Delta\Omega_c$ change in the carrier wave frequency results in a 
$\Delta \Omega_0=\mp\Delta\Omega_c$ change in the locked laser frequency.

\subsection{ESB vs DSB}\label{sec:ESBDSB}
\begin{figure}[h]
    \centering
    \subfloat[\label{fig:standardPDH}]{
        \includegraphics[width=0.47\textwidth]{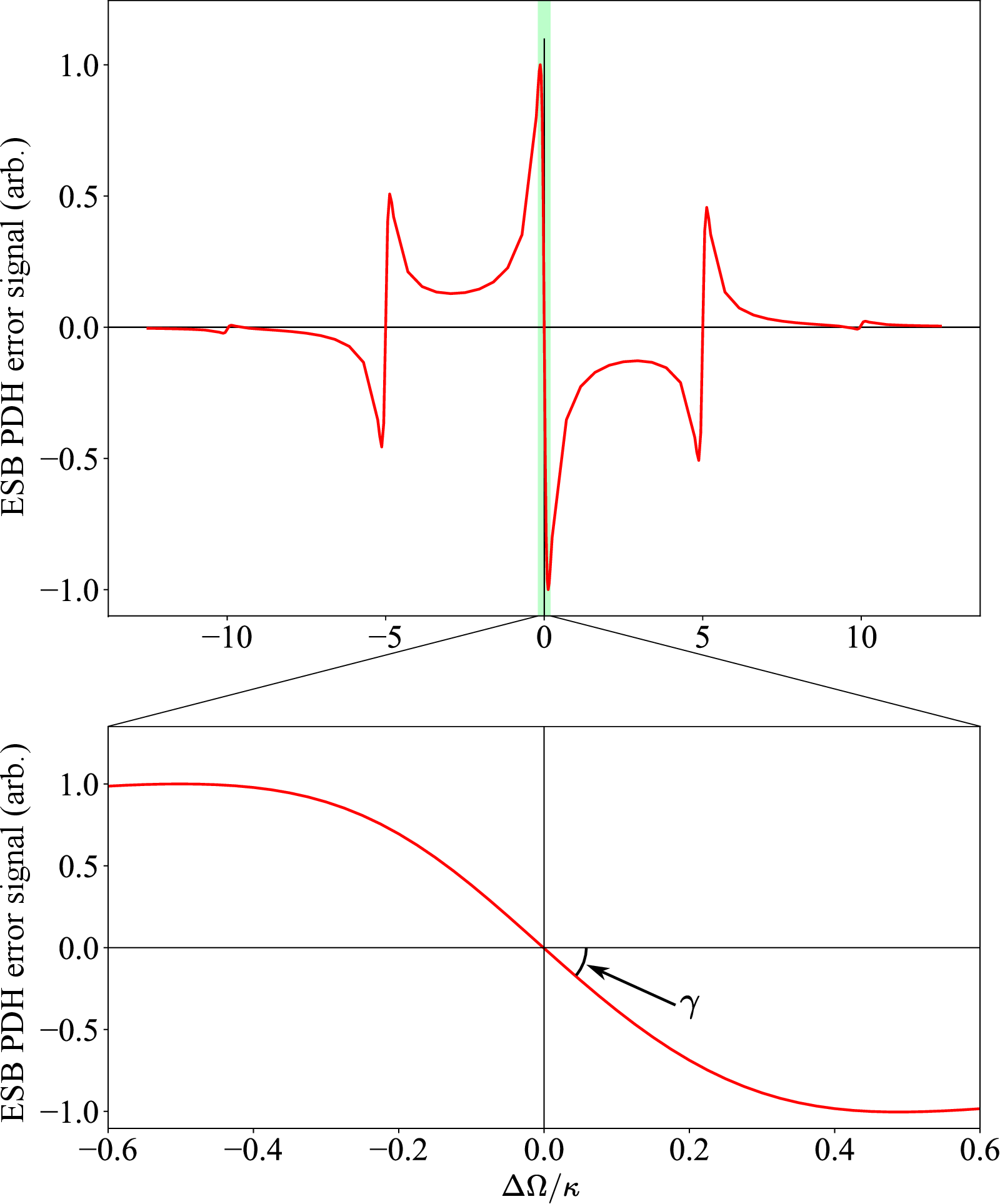}
    }\hfill
    \subfloat[\label{fig:ESB-DSB}]{
        \includegraphics[width=0.47\textwidth]{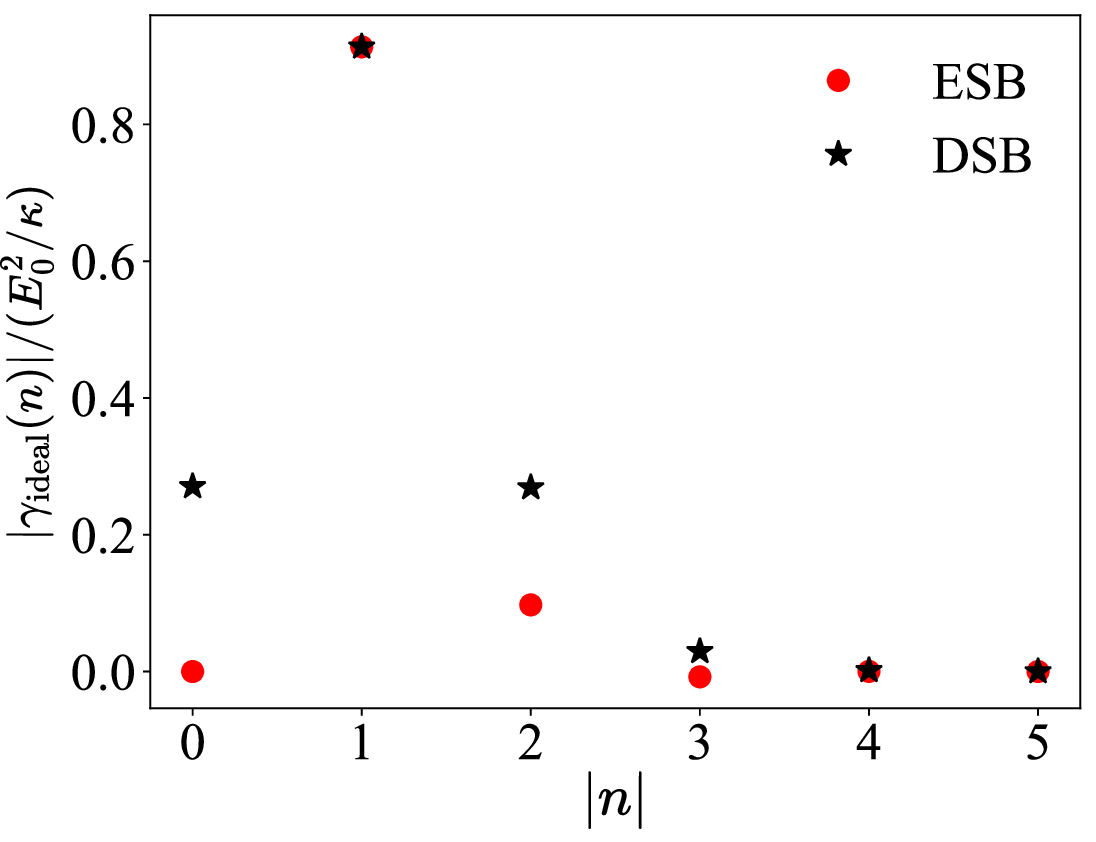}
    }
    \caption{(a) Error signal near cavity resonance; $\gamma$ associated with the error signal is illustrated in the zoomed-in plot. (b) $\left|\gamma_\textrm{ideal}\right|$ as a function of spectral triplet of order $n$ for the ideal ESB and DSB locking schemes at $\beta_c=1.84$ and $\beta_m=1.01$.
    }
    \label{fig:ESB-DSB-gamma}
\end{figure}

As mentioned in the Introduction, a key advantage of ESB locking over DSB locking is that ESB produces a less complicated optical spectrum with strongly suppressed spurious sidebands, which reduces the likelihood of unintended lock points, simplifies lock acquisition, and doubles the tuning range relative to DSB~\cite{Thorpe:08,Livas_2009}. To quantify this advantage, we evaluate the PDH-like error signals associated with each spectral triplet consisting of a central component at $\Omega_0+n\Omega_c$ and surrounding sidebands at $\Omega_0+n\Omega_c\pm\Omega_m$ in the vicinity of a cavity resonance.

We refer to the set of three fields at frequencies $\Omega_0+n\Omega_c-\Omega_m,\Omega_0+n\Omega_c,\Omega_0+n\Omega_c+\Omega_m$ as the spectral triplet of order $n$. When the central component $\Omega_0+n\Omega_c$ lies near a cavity resonance, this triplet generates a PDH error signal upon demodulation at $\Omega_m$. Only the $|n|=1$ triplet corresponds to the intended locking feature in ESB/DSB; higher-order triplets ($|n|\neq 1$) generate additional spurious PDH error signals and can therefore create unintended lock points if their slopes and, therefore, the error signal size, are sufficiently large.

The error-signal gain $\gamma$ is defined as the slope of the demodulated error signal in the vicinity of the cavity resonance, as illustrated in Fig.~\ref{fig:standardPDH}. In Fig.~\ref{fig:ESB-DSB}, we compare the magnitude of the ideal $\gamma(n)$ for ESB and DSB. The ideal error-signal gain for ESB arising from the spectral triplet at ${\Omega_0+n\Omega_c,\ \Omega_0+n\Omega_c\pm\Omega_m}$ is
\begin{equation}\label{eq:ESB_ESG}
\gamma_{\mathrm{ideal,ESB}}(n)
=-\frac{8E_0^2}{\kappa}J_n^2(\beta_c)J_0(n\beta_m)J_1(n\beta_m),
\end{equation}
whereas for DSB it is
\begin{equation}\label{eq:DSB_ESG}
\gamma_{\mathrm{ideal,DSB}}(n)
=-\frac{8E_0^2}{\kappa}J_n^2(\beta_c)J_0(\beta_m)J_1(\beta_m),
\end{equation} where $\kappa$ is the cavity linewidth.

These expressions show that, for $\beta_m=\beta_{\mathrm{opt}}=1.01~\mathrm{rad}$ (the optimal modulation index for both schemes~\cite{Thorpe:08}), the spurious PDH discriminators with $|n|\neq 1$ are either absent ($n=0$) or more strongly suppressed ($|n|\ge 2$) in ESB than in DSB. In particular, the absence of the $n=0$ discriminator eliminates the corresponding unintended lock point, which effectively doubles the usable tuning range for ESB relative to DSB.

\subsection{QAM-based ESB signal generation}\label{sec:QAM}
\begin{figure}[tb!]
\includegraphics[width=\linewidth]{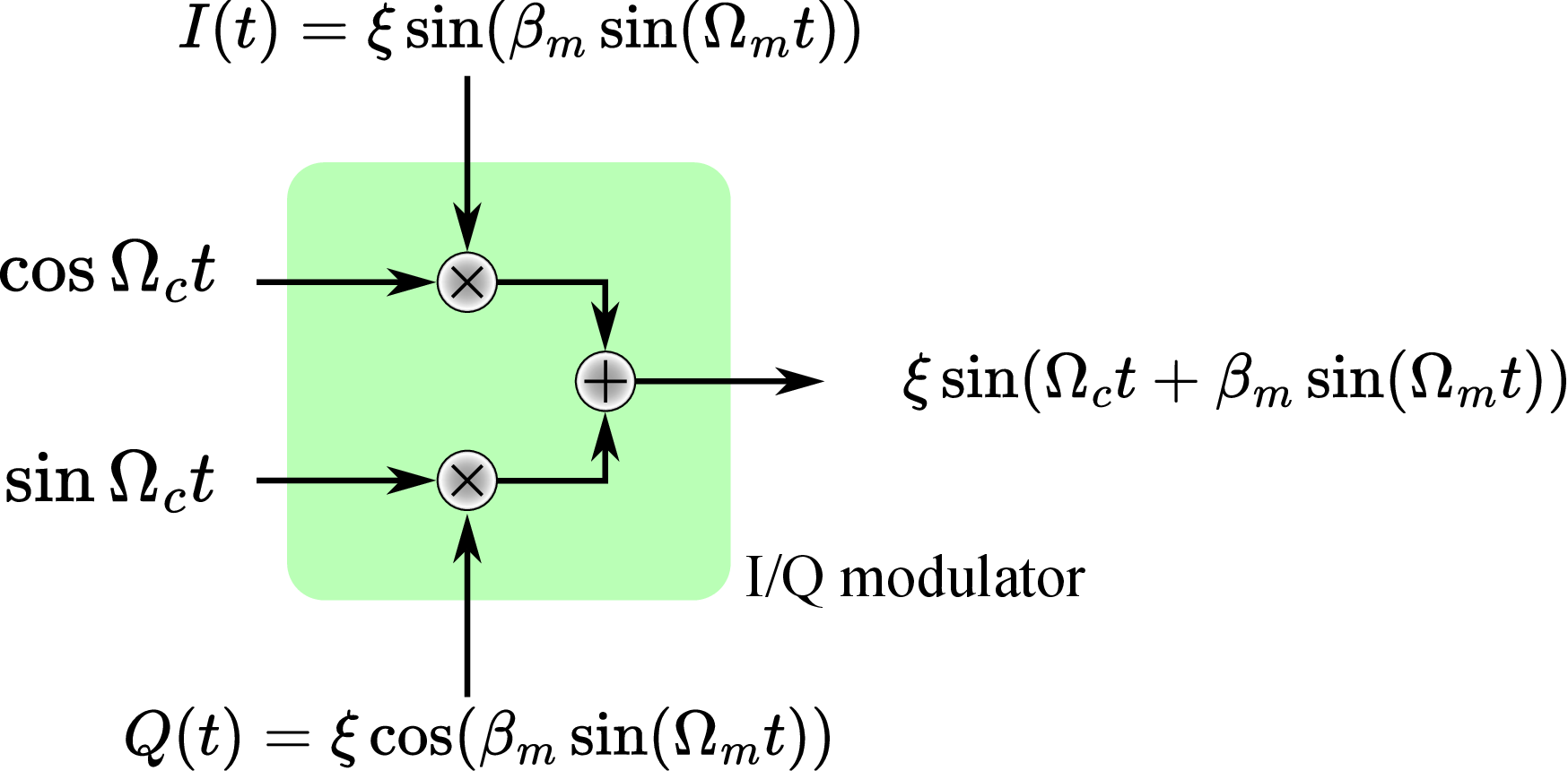}
\caption{\label{fig:FM}
QAM scheme: Carrier wave signals $\cos (\Omega_c t)$ and $\sin (\Omega_c t)$ are mixed with baseband signals $I(t)$ and $Q(t)$ and summed. 
}
\label{fig:QAM}
\end{figure}

Expressing the ideal phase-modulated rf signal $V_{\textrm{EOM}}(t)$ for ESB in terms of its quadrature components
\begin{align}
    V_{\text{EOM}}(t)&= \xi\sin[\Omega_{c}t+\beta_{m}\sin(\Omega_{m}t)]\nonumber\\
    &=\xi\sin[\beta_{m}\sin(\Omega_{m}t)]\cos(\Omega_{c}t) \nonumber \\ 
    &\ \ \ \ +\xi\cos[\beta_{m}\sin(\Omega_{m}t)] \sin(\Omega_{c}t) \nonumber \\
    &\equiv I(t) \cos(\Omega_{c}t) + Q(t) \sin(\Omega_{c}t) \label{eq:QAM},
\end{align}
motivates the use of QAM for constructing rf signals. 
Here $\xi$ is the amplitude of $V_{\textrm{EOM}}(t)$. The in-phase and quadrature-phase baseband channels $I(t)$ and $Q(t)$ amplitude modulate two carrier waves that are $\pi/2$ radians out of phase, which then sum to $V_{\textrm{EOM}}(t)$ (see Fig.~\ref{fig:QAM}).
However, inevitable nonidealities in QAM---called I/Q impairments---distort the generated phase-modulated rf signal, which affects the locked laser frequency spectrum.

\subsection{\label{sec:theory1}I/Q impairments and their effects}

\begin{figure}[h]
    \includegraphics[width=0.96\linewidth]{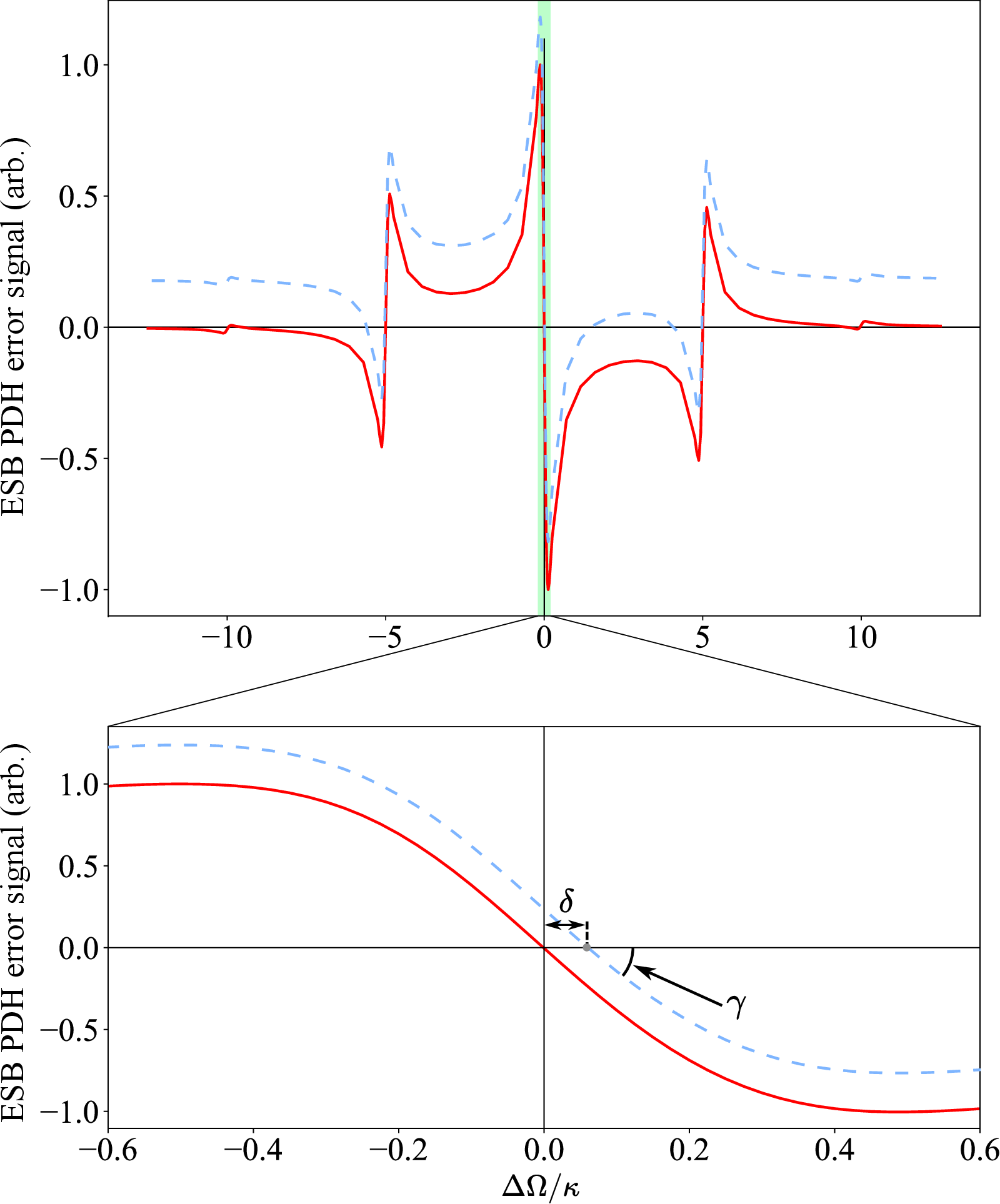}
    \caption{Cartoon of ESB error signals near cavity resonance: The solid (dashed) line indicates the ESB error signal without (with) impairments. The gain $\gamma$ and offset $\delta$ of the ESB error signal with impairments in the vicinity of the cavity resonance is shown in the zoomed-in plot.
    }\label{fig:perfect}
\end{figure}

\begin{figure}[h]
    \centering
    \subfloat[\label{fig:IQmagnitudeError}]{
        \includegraphics[width=0.95\linewidth]{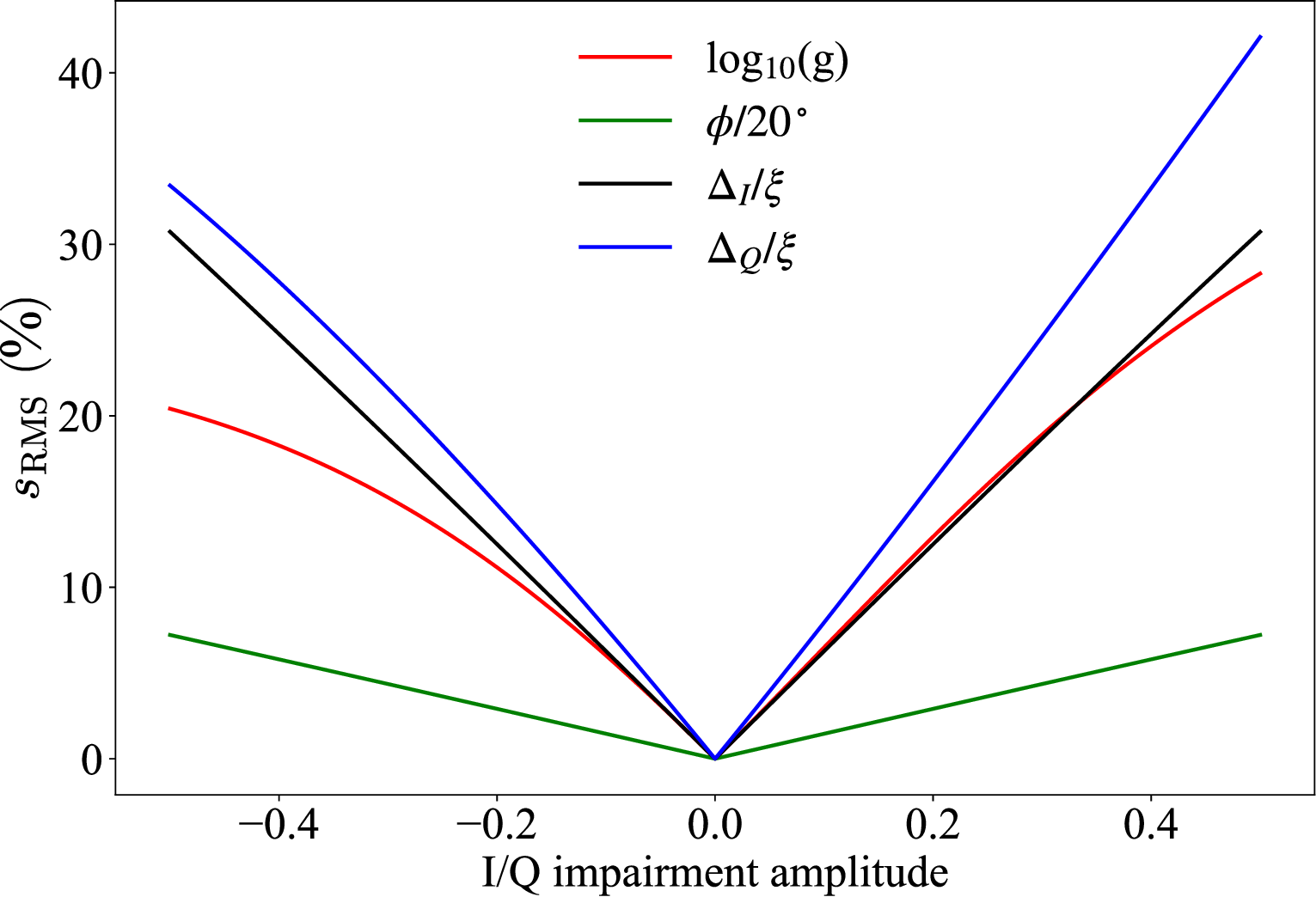}
    }\hfill
    \subfloat[\label{fig:IQphaseError}]{
        \includegraphics[width=0.95\linewidth]{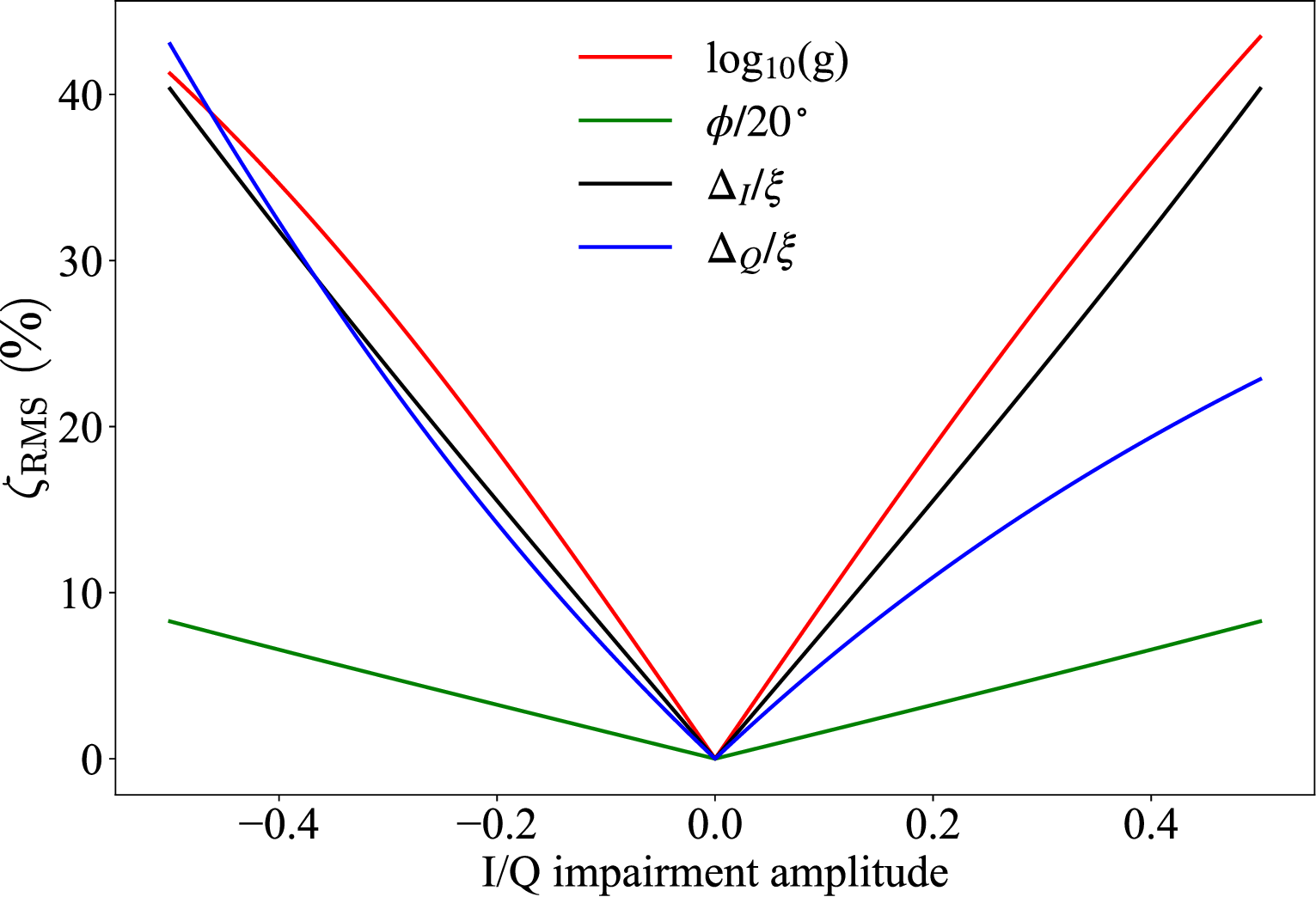}
    }\hfill
    \caption{Effect of the I/Q impairments on the (a) I/Q magnitude error $s_{\mathrm{RMS}}(q)$ as a function of the I/Q impairment amplitudes: $\log _{10}(\mathrm{g}), \phi / 20^{\circ}, \Delta_I / \xi, \Delta_Q / \xi$, and (b) I/Q phase error $\zeta_{\mathrm{RMS}}(q)$ as a function of the I/Q impairment amplitudes: $\log _{10}(\mathrm{g}), \phi / 20^{\circ}, \Delta_I / \xi, \Delta_Q / \xi$. 
    }\label{fig:IQErrors}
\end{figure}

 \begin{figure}[h]
  \subfloat[\label{fig:ESG}]{
        \includegraphics[width=0.96\linewidth]{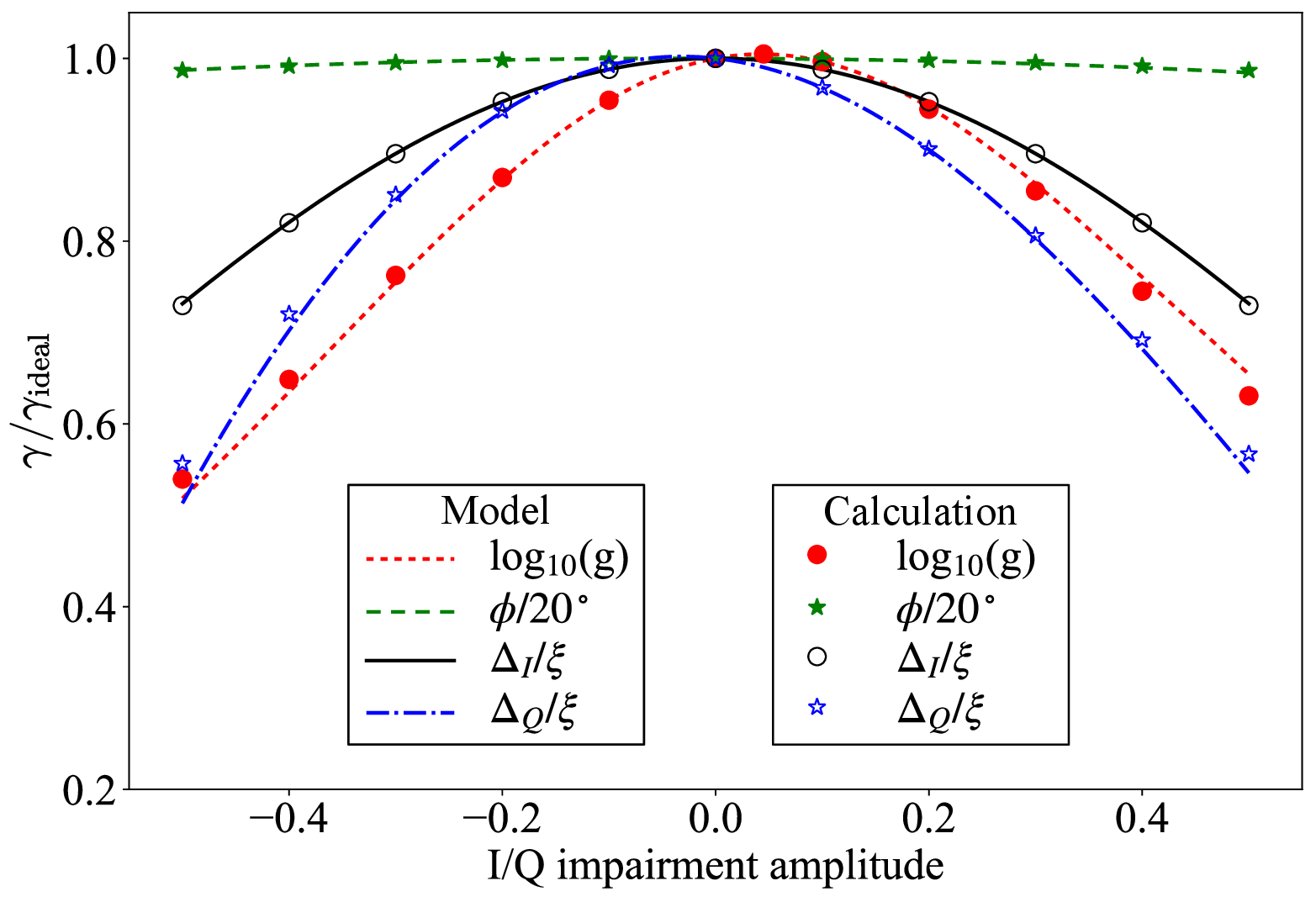}
    }\hfill
    \subfloat[\label{fig:ESO}]{
        \includegraphics[width=0.96\linewidth]{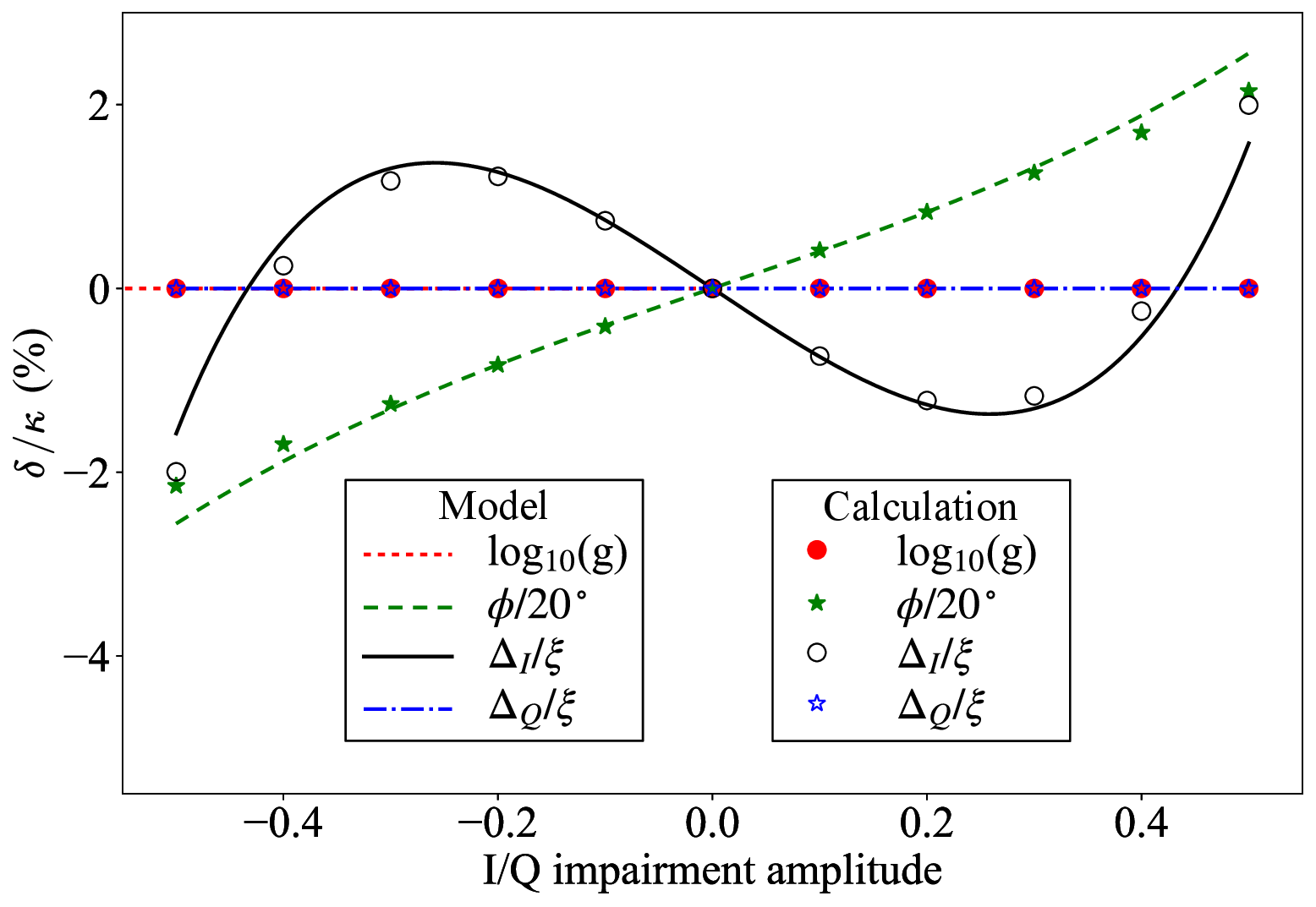}
        }\hfill
    \caption{
    Effect of the I/Q impairments on the ESB error signal: (a)  $\gamma/\gamma_{\textrm{ideal}}$ as a function of the I/Q impairment amplitudes: $\log _{10}(\mathrm{g}), \phi / 20^{\circ}, \Delta_I / \xi, \Delta_Q / \xi$, (b) $\delta$ / $\kappa$ as a function of the I/Q impairment amplitudes: $\log_{10}(\mathrm{g}), \phi / 20^{\circ}, \Delta_I / \xi, \Delta_Q / \xi$. Traces (points) represent analytically (numerically) computed quantities.
   }\label{fig:esberrors}
\end{figure}

I/Q impairments~\cite{1275708,4389078,Ghannouchi2013,Mohammadian2021} modify the ideal phase-modulated signal $V_{\text{EOM}}(t)$ in Eq.~\eqref{eq:QAM} to
\begin{align}
    V_{\text{EOM}}(t)&= I(t) \cos(\Omega_{c}t) + Q(t) \sin(\Omega_{c}t) \label{eq:badQAM}\\
    &=\sqrt{I(t)^2+Q(t)^2}\sin\left(\arctan\frac{I(t)}{Q(t)}+\Omega_ct\right),
\end{align}
where 
\begin{equation}
\begin{aligned}
I(t) &= \xi_I \sin[\beta_{m}\sin(\Omega_{m}t)] + \Delta_I, \\
Q(t) &= \xi_Q \cos[\beta_{m}\sin(\Omega_{m}t)+\phi] + \Delta_Q.
\end{aligned}\label{eq:trueQAM}
\end{equation}
$\phi$ is the phase imbalance between the I/Q channels, $\Delta_I$ and $\Delta_Q$ are the DC offsets,
and
\begin{align*}
\xi_I &= g\xi \sqrt{\frac{2}{1+g^2}} & {\rm and} && \xi_Q=\xi\sqrt{\frac{2}{1+g^2}}
\end{align*}
are the amplitudes expressed in terms of the gain imbalance $g={\xi_I}/{\xi_Q}$ and $\xi=\sqrt{(\xi^2_I+\xi^2_Q)/2}$.

Drifts in the I/Q impairments translate into drifts of the ESB error signal. The primary error-signal parameters affected are the gain $\gamma$ and the error-signal offset $\delta$ (see Fig.~\ref{fig:perfect}). For an ideal error signal, $\delta=0$ and $\gamma=\gamma_{\mathrm{ideal}}$ (see the previous section). In practice, I/Q impairments modify both $\gamma$ and $\delta$. These hardware impairments can be compensated by appropriately engineering the baseband signals. In particular, one can adjust the parameters $g$, $\phi$, $\Delta_I$, and $\Delta_Q$ of the I- and Q-channel waveforms so that the resulting modulation approaches the ideal phase-modulation case as closely as possible [Eq.~\eqref{eq:QAM}].

The quality of a phase-modulated rf signal is typically assessed with a spectrum analyzer by extracting the I/Q phasor and reporting the RMS I/Q magnitude error $s_{\mathrm{RMS}}(q)$ and RMS I/Q phase error $\zeta_{\mathrm{RMS}}(q)$,
\begin{align}
s_{\mathrm{RMS}}(q) &= \sqrt{\frac{1}{T}\int_{0}^{T} s(q;t)^{2}\,dt}, \\
\zeta_{\mathrm{RMS}}(q) &= \sqrt{\frac{1}{T}\int_{0}^{T} \zeta(q;t)^{2}\,dt},
\end{align}
where $s(q;t)$ is the instantaneous I/Q magnitude error, $\zeta(q;t)$ is the instantaneous I/Q phase error and  $q \in \left\{\log_{10}(g),\ \phi/20^{\circ},\ \Delta_{I}/\xi,\ \Delta_{Q}/\xi\right\}$. We define
\begin{align} s(q;t)&=\sqrt{\frac{I(q;t)^2+Q(q;t)^2}{I(q=q_0;t)^2+Q(q=q_0;t)^2}}-1\nonumber\\ &=\frac{1}{\xi}\sqrt{I(q;t)^2+Q(q;t)^2}-1\\ \zeta(q;t)&=\frac{1}{\beta_m}\left(\arctan\frac{I(q;t)}{Q(q;t)}-\arctan\frac{I(q=q_0;t)}{Q(q=q_0;t)}\right)\nonumber\\ &=\frac{1}{\beta_m}\arctan\frac{I(q;t)}{Q(q;t)}-\sin(\Omega_mt) \end{align}
where $q_{0}$ denotes the ideal (impairment-free) setting and $\xi \equiv \sqrt{I(q_{0};t)^{2}+Q(q_{0};t)^{2}}$ is the corresponding I/Q magnitude (constant in time for the ideal waveform). We define $q$ in normalized dimensionless units so that the effects of the various I/Q impairments on the ESB error signal can be quantified on an equal footing.

Fig.~\ref{fig:IQErrors} shows the calculated RMS I/Q errors as a function of the dimensionless impairment amplitude. For small impairments, $s_{\mathrm{RMS}}(q)$ and $\zeta_{\mathrm{RMS}}(q)$ scale approximately linearly with $\log_{10}(g)$, $\phi/20^{\circ}$, $\Delta_{I}/\xi$, and $\Delta_{Q}/\xi$
, providing a direct method for estimating embedded I/Q impairments from measured I/Q errors of the modulated rf signal.  

We analytically and numerically study the effect of I/Q impairments on the ESB error signal, as shown in Fig.~\ref{fig:esberrors} (see Appendix~\ref{sec:closedformequations} for details). Solid curves indicate the analytical model, and points indicate numerical calculations. The analytical model is obtained from a second-order Taylor expansion of the error signal. The dependence of the normalized error-signal parameters, $\gamma/\gamma_{\textrm{ideal}}$ and $\delta/\kappa$, on each I/Q impairment amplitude---$\log_{10}(g)$, $\phi/20^{\circ}$, $\Delta_I/\xi$, and $\Delta_Q/\xi$---is shown in Fig.~\ref{fig:ESG} and Fig.~\ref{fig:ESO}, respectively. The calculations were performed for $\Omega_m/(2\pi)=10~\mathrm{MHz}$ and $\kappa/(2\pi)=20~\mathrm{kHz}$, and for the theoretically optimal ESB parameters $\beta_m=\beta_{\textrm{opt}}=1.01~\mathrm{rad}$ and $\beta_c=1.84~\mathrm{rad}$~\cite{Thorpe:08}. Here $\gamma_{\textrm{ideal}}$ is the ESB error-signal gain in the absence of I/Q impairments, and $\kappa$ is the reference-cavity linewidth. The curves (points) in Fig.~\ref{fig:ESG} and Fig.~\ref{fig:ESO} correspond to the analytical (numerical) results.

As shown in Fig.~\ref{fig:ESG}, $\gamma$ is first-order insensitive to fluctuations in the I/Q impairment amplitudes, which helps mitigate gain-margin sensitivity in the laser-frequency stabilization servo. By contrast, $\delta/\kappa$ is first-order sensitive to $\phi$ and $\Delta_I/\xi$, with slopes of $-0.22~\%/^{\circ}$ and $7.8~\%$ per unit of $\Delta_I/\xi$, respectively, but is identically zero for all values of $\log_{10}(g)$ and $\Delta_Q/\xi$ (see Fig.~\ref{fig:ESO}). A nonzero $\delta$ directly translates into a frequency offset of the locked laser (see Eq.~\eqref{eq:offset}), so $\phi$ and $\Delta_I/\xi$ must be sufficiently stable to suppress long-term frequency drifts.

Because both the measurable RMS I/Q errors $s_{\mathrm{RMS}}(q)$ and $\zeta_{\mathrm{RMS}}(q)$ (Fig.~\ref{fig:IQErrors}) and the ESB frequency offset $\delta(q)$ (Fig.~\ref{fig:esberrors}) scale linearly with a single small I/Q impairment amplitude $q$, the measured I/Q errors can be mapped directly to an estimated frequency offset using
\begin{align}
\frac{\delta(q)}{\mathrm{RMS\ IQ\ Error}(q)}
=\frac{\delta(q)}{q}\times\frac{q}{\mathrm{RMS\ IQ\ Error}(q)}.
\end{align}
From Fig.~\ref{fig:esberrors}, we find that $\phi$ and $\Delta_I$ dominate the contributions to $\delta$, with $\delta(\phi)/\phi=-0.0022~\kappa/^{\circ}$ and $\delta(\Delta_I)/\Delta_I=0.078~\kappa/\xi$. Combining these slopes with the corresponding $q/s_{\mathrm{RMS}}(q)$ and $q/\zeta_{\mathrm{RMS}}(q)$ slopes obtained from Fig.~\ref{fig:IQErrors} yields
\begin{align}
\frac{\delta(\phi)}{s_{\mathrm{RMS}}(\phi)}&=0.30~\kappa, \label{eq:ESO_sphi}\\
\frac{\delta(\phi)}{\zeta_{\mathrm{RMS}}(\phi)}&=0.27~\kappa, \label{eq:ESO_zetaphi}\\
\frac{\delta(\Delta_I)}{s_{\mathrm{RMS}}(\Delta_I)}&=0.12~\kappa, \label{eq:ESO_sI}\\
\frac{\delta(\Delta_I)}{\zeta_{\mathrm{RMS}}(\Delta_I)}&=0.10~\kappa. \label{eq:ESO_zetaI}
\end{align}
Although these relations are derived under the assumption that only a single I/Q
impairment is present at a time, they provide a useful diagnostic upper bound on
ESB-induced frequency offsets. For an optical cavity linewidth of $\kappa =
2\pi \times 20~\mathrm{kHz}$, a $0.3\%$ RMS I/Q magnitude error $s_{\mathrm{RMS}}
$ implies a maximum frequency offset of $\delta/(2\pi) \approx 18~\mathrm{Hz}$,
while a $0.3\%$ RMS I/Q phase error $\zeta_{\mathrm{RMS}}$ implies $\delta/(2\pi)
\approx 16~\mathrm{Hz}$. Drifts in the I/Q impairment amplitudes therefore
translate directly into drifts of the locked frequency offset. Such offsets are
significant when interrogating ultranarrow optical clock transitions, whose
natural linewidths are typically in the millihertz to tens-of-millihertz range~\cite{Rosenband2007,Porsev2004,Muniz2021,Dolde2025,Nicholson2015,Takamoto2003,Derevianko2011}.

\subsection{Other practical considerations}
\label{sec:passband}
Passband bandwidth limitations in the quadrature modulator and baseband signal chain can distort the generated phase-modulated rf waveform, particularly when the phase-modulation index $\beta_m$ and modulation frequency $\Omega_m$ are large. This is relevant for ESB locking because the theoretically optimal modulation index, $\beta_m=\beta_{\mathrm{opt}}=1.01~\mathrm{rad}$, is relatively large. In addition, large modulation frequencies $\Omega_m$ (typically in the HF range) are desirable in PDH locking to improve signal-to-noise ratio by operating above dominant $1/f$ technical noise.

In standard PDH locking, the EOM itself defines the relevant passband, which is typically in the super high frequency (SHF) range; consequently, passband limitations are usually negligible for the PDH error signal. This is not necessarily true for ESB locking, where the effective passband is often set by the quadrature modulator baseband bandwidth and associated analog front-end, which may be comparable to (rather than far above) the desired $\Omega_m$.

A limited passband can truncate higher-order sidebands generated during phase modulation. Moreover, gain ripple and phase dispersion across the passband distort the amplitude and phase relationships among the lower-order sidebands. These distortions appear as I/Q magnitude and phase errors and can lead to parasitic frequency offsets. This is similar to residual amplitude modulation (RAM) which is known to be detrimental in PDH locking~\cite{Thorpe:08,Shi2018,Gillot2022} because it manifests as a DC offset in the demodulated error signal that can drift in time. For this reason, a wide passband with low gain ripple and minimal phase dispersion is desirable.

Finally, the carrier wave oscillator must be chosen with appropriate phase noise performance, long-term stability, and wideband tunability. Phase noise on the carrier wave oscillator limits the ultimate stabilized laser linewidth and carrier-frequency drift maps directly onto drift of the locked laser frequency. In addition, unfavorable transients during the tuning of the carrier wave oscillator frequency can break the ESB lock and thereby limit the practical tuning range. See Appendix~\ref{sec:carrierwave} for more details on the carrier-wave oscillator.

\begin{widetext}
\begin{figure}[!tb]
\centering
\includegraphics[width=0.9\textwidth]{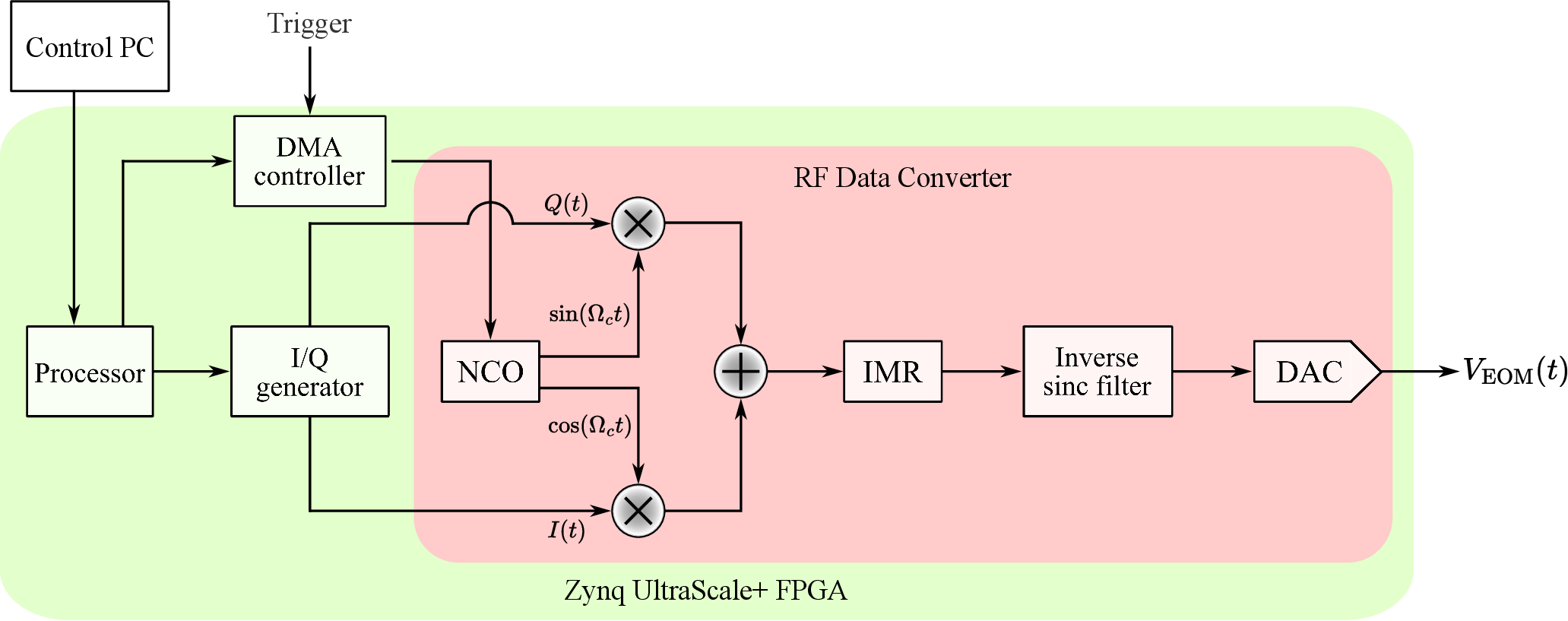}
\begin{minipage}{\textwidth}
\caption{\label{fig:blockdiagram}Block diagram of the gateware layout for the RFSoC-based ESB rf signal ($V_{\textrm{EOM}}(t)$) generation architecture. The full forms of the abbreviations used in the figure are listed as follows: DMA -- Direct memory access, NCO -- Numerically controlled oscillator, IMR -- Image rejection filter.
}
\end{minipage}
\end{figure}
\end{widetext}
\section{RFSoC-based device architecture\label{sec:design}}
We design and implement a direct SDR~\cite{booksdr} to synthesize the phase-modulated rf signal $V_{\textrm{EOM}}(t)$ defined in Eq.~\eqref{eq:QAM}. The complete digital signal generation, illustrated in Fig.~\ref{fig:blockdiagram}, is implemented on an AMD Zynq UltraScale+ RFSoC ZU48DR FPGA. The process begins with a control PC that supplies the fundamental QAM parameters ($\Omega_m, \Omega_c, \beta_m$) and compensation parameters ($\phi', \xi_I', \xi_Q', \Delta_I', \Delta_Q'$) to the on-chip processing system (PS). Based on these inputs, a digital I/Q generator block, comprising a phase accumulator, trigonometric phase-to-amplitude look-up tables, and interpolation modules, synthesizes the baseband waveforms $I(t)$ and $Q(t)$. As detailed in Eq.~\eqref{eq:trueQAM} and Sec.~\ref{sec:performance}, applying these compensation parameters digitally pre-distorts the baseband signals to cancel the system's intrinsic I/Q impairments ($\xi_I, \xi_Q, \Delta_I, \Delta_Q, \phi$), thereby minimizing the RMS I/Q errors.

For frequency up-conversion, these conditioned baseband signals are routed to the FPGA's RF Data Converter (RFDC) hardware core~\cite{rfdc2024}. This core integrates a numerically controlled oscillator (NCO), a digital up-converter, image rejection (IMR) and inverse sinc filters, and high-bandwidth RF-DACs into a unified hardware module. Operating at a sample rate $f_s$ of $9.8304$ gigasamples per second (GSPS), the RFDC synthesizes a clean modulated output across a wide bandwidth extending from DC to $0.2 f_s$~\cite{rfdc2024}. Finally, the internal RF-DACs convert this up-converted digital waveform into the analog rf drive signal for the EOM.

The carrier frequency generated by the NCO can be updated in real time via a custom direct memory access (DMA) controller. By rapidly updating the frequency tuning word (FTW), the device enables wideband carrier tuning while maintaining phase continuity in the rf output. As a result, the laser remains locked while the carrier frequency is swept over a broad range. We validate the smoothness and stability of this RFSoC-based tuning through electronic and optical measurements in the subsequent sections.

We also designed an ESB rf signal generator based on Analog Devices' ADALM-PLUTO evaluation board. The device uses a phase-locked loop (PLL) circuit with integrated voltage-controlled oscillators (VCOs) to generate the carrier wave (see Appendix~\ref{sec:carrierwave}).\footnote{The details of the ADALM-PLUTO-based design can be found in its GitHub repository: \url{https://github.com/JQIamo/Electronic_Sideband_Locking_Pluto}.}

\section{Experimental characterizations\label{sec:performance}}
\begin{figure}[h]
\centering
    \subfloat[\label{fig:iqplot_RFSoC} ]{
        \includegraphics[width=0.9\linewidth]{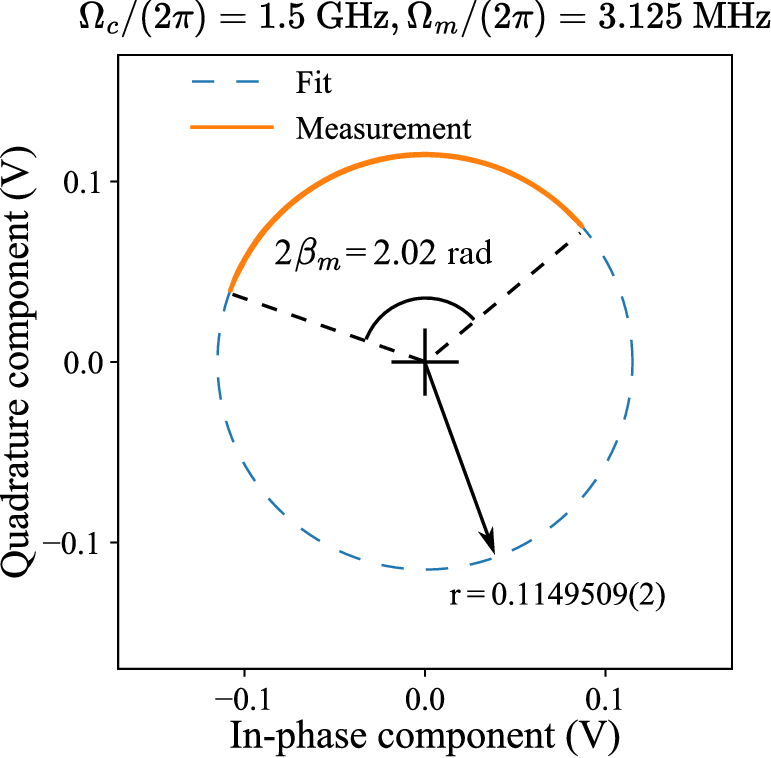}}\hfill
    \subfloat[\label{fig:RFSoC}]{
        \includegraphics[width=0.85\linewidth]{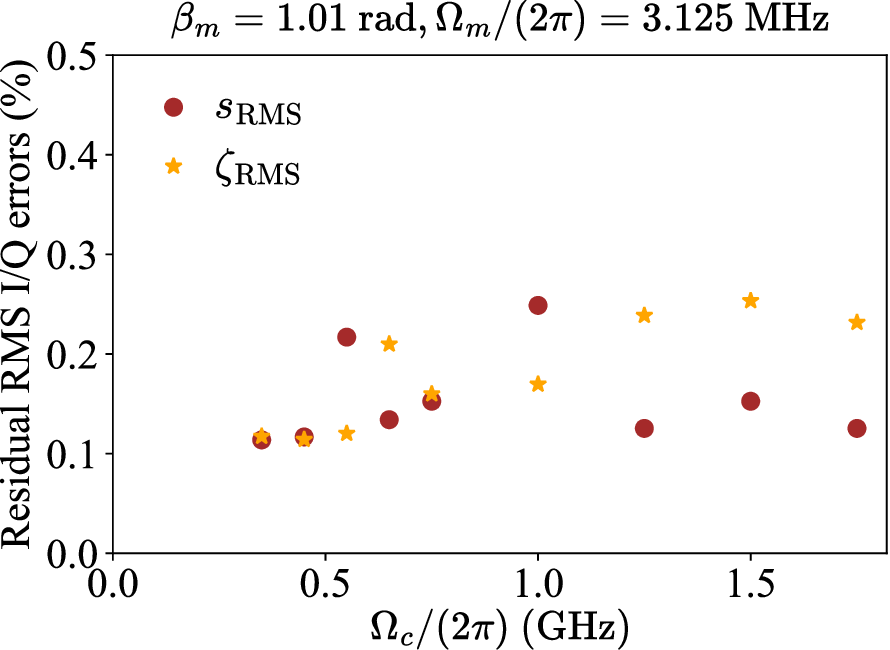}}
    \caption{\label{fig:residual2}
Measured QAM quality at $\beta_m=\beta_{\textrm{opt}}$ and $\Omega_m/(2\pi)=3.125$ MHz.
(a) Measured I/Q phasor trajectory plotted with a circular fit circle.
(b) Residual RMS I/Q magnitude error and residual RMS I/Q phase error as a function of $\Omega_{c}/(2\pi)$.}
\end{figure}

\subsection{Electronic characteristics\label{sec:Electronic Characteristics}}

We follow the standard procedure for quantifying the quality of QAM: measure the I/Q phasor (a constellation diagram-type measurement); measure the RMS I/Q magnitude error $s_\textrm{RMS}$ and RMS I/Q phase error $\zeta_\textrm{RMS}$ (an error vector magnitude-type measurement)~\cite{Acar1,Witte2001,Razavi2011,Nash2009,Technologies,Denisowski2022,NuWaves2019}. The I/Q phasor for an ideal phase-modulated signal lies on a circle (see Sec.~\ref{sec:theory1} for more details). Deviations of the measured I/Q phasor from this reference circle indicate the type and strength of I/Q impairments. We use a Rohde \& Schwarz FSV spectrum analyzer for these measurements. 

The spectrum analyzer directly measures the I/Q phasor and the RMS I/Q magnitude error; however, we extract the I/Q phase error by fitting the angle of the measured I/Q phasor to that of an ideal phase-modulated I/Q phasor: $\beta_m\sin(\Omega_m t)$. The root mean square of the fit residuals normalized by $\beta_m$ yields the RMS I/Q phase error. 
We compensate for the I/Q impairments in the hardware by minimizing the RMS I/Q magnitude error via manually tuning $\xi_I'$, $\xi_Q'$, $\Delta_I'$, $\Delta_Q'$, and $\phi'$. We verified that there is no compensation needed for the design, which is reasonable as the entire I/Q signal generation and modulation are performed in the digital domain.

First, we measure the I/Q phasor and fit it to a circle. 
The measured I/Q phasor at $\beta_m=\beta_{\textrm{opt}}$, $\Omega_{m}/(2\pi)=3.125$ MHz, $\Omega_{c}/(2\pi)=1.5$ GHz for the RFSoC-based design is shown in Fig.~\ref{fig:iqplot_RFSoC}. The radius of the fitted circle is $0.1149509(2)$ V and its center is located at $[-0.00002497(8),0.0000536(3)]$ V.\footnote{The numbers in the brackets indicate the standard errors.} 
The fit quality of the measured I/Q phasor to a circle reflects the quality of the phase modulation. Next, we study the behavior of residual I/Q phase and magnitude errors as a function of the carrier wave frequency $\Omega_c$. We set $\beta_m=\beta_{\textrm{opt}}$ and $\Omega_{m}/(2\pi)=3.125$ MHz for the measurement. We measured the I/Q errors across a carrier wave frequency range from $350$ MHz to $1.75$ GHz. Fig.~\ref{fig:RFSoC} shows the measured residual RMS I/Q errors as a function of $\Omega_c$. We measure an RMS I/Q magnitude error and an RMS I/Q phase error less than $0.3$~\% throughout the carrier wave frequency range. 

\begin{figure}[h!]
    \centering
    \subfloat[\label{fig:ESBL_jump}]{
        \includegraphics[width=0.8\linewidth]{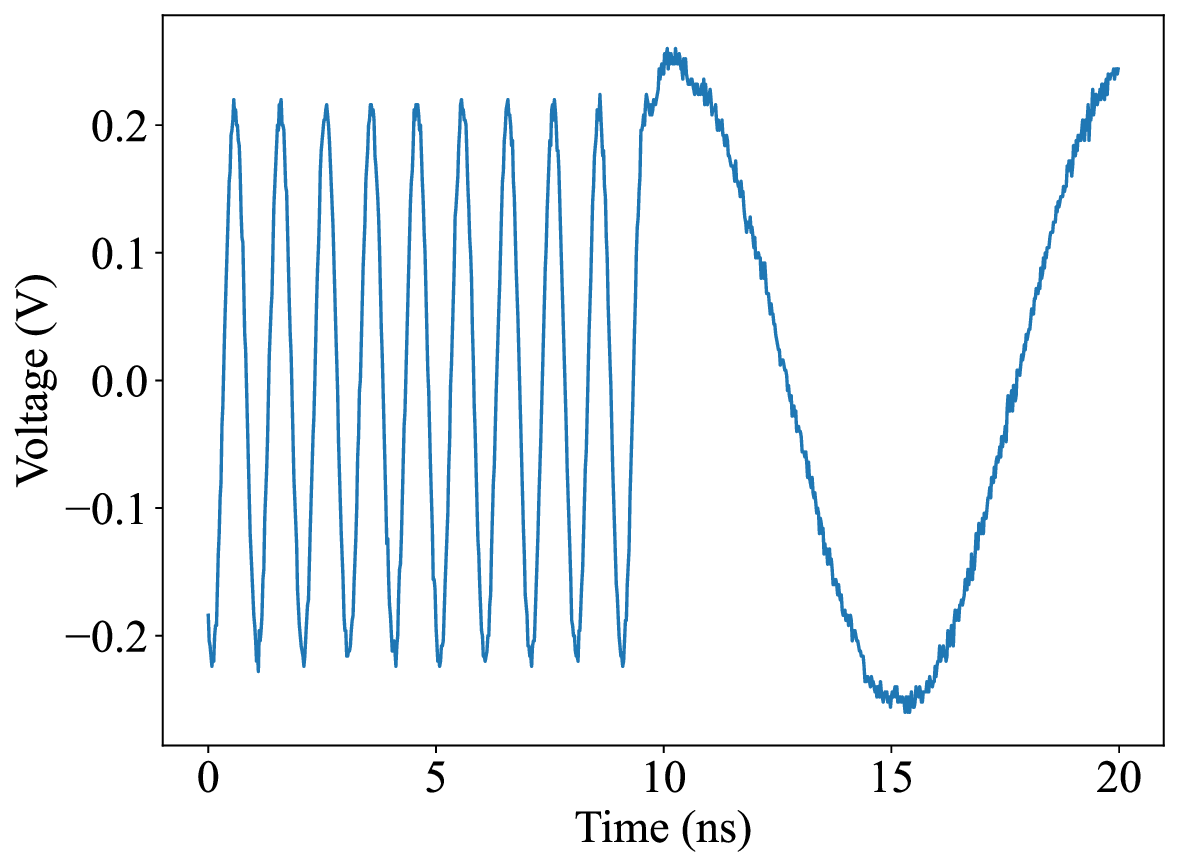}
    }\hfill
    \subfloat[\label{fig:ESBL_ramp}]{
        \includegraphics[width=0.8\linewidth]{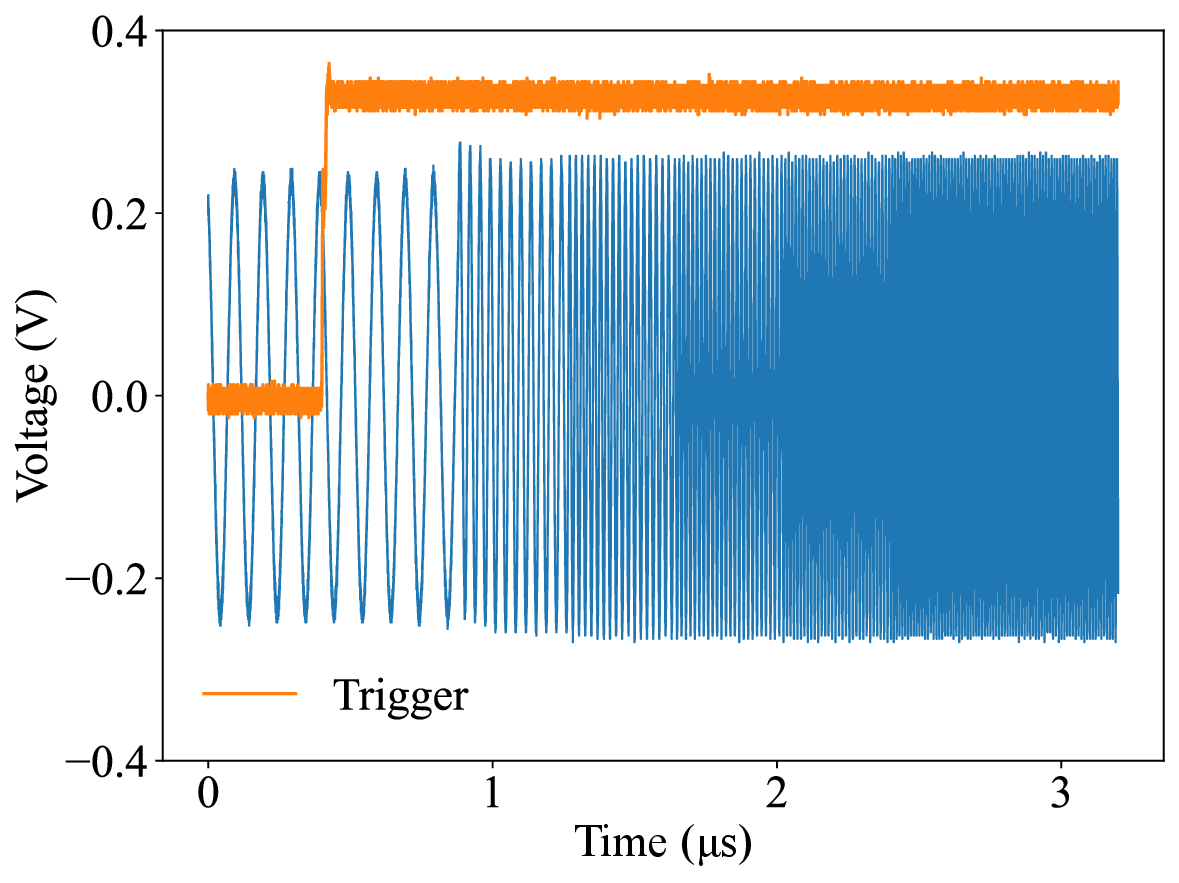}
    }\hfill
    \caption{Carrier output during NCO tuning: (a) frequency jump from 1 GHz to 100 MHz. (b) Programmed frequency ramp from 10 MHz to 100 MHz in 5 steps.}
    \label{fig:freqjumps}
\end{figure}

Moreover, we characterize the RFSoC’s NCO tuning behavior by measuring the RF-DAC output during programmed frequency updates. Fig.~\ref{fig:freqjumps} shows oscilloscope traces (50~GS/s) of the carrier waveform during an abrupt frequency change (see Fig.~\ref{fig:ESBL_jump}) and a stepwise frequency ramp (see Fig.~\ref{fig:ESBL_ramp}). In Fig.~\ref{fig:ESBL_jump}, we update the NCO frequency tuning word (FTW) to change the carrier from $1~\mathrm{GHz}$ to $100~\mathrm{MHz}$; the transition is glitch-free at the oscilloscope resolution, with no observable phase discontinuity in the output waveform. In Fig.~\ref{fig:ESBL_ramp}, we stream a list of five FTWs via DMA to ramp the carrier from $10~\mathrm{MHz}$ to $100~\mathrm{MHz}$ in discrete steps. The waveform remains phase-continuous across each update, and the time spacing between consecutive steps yields an estimated FTW update period of approximately $380~\mathrm{ns}$ for DMA-driven tuning.

\subsection{Optical characteristics}\label{sec:optical}
\begin{figure}[h]
    \includegraphics[width=0.96\linewidth]{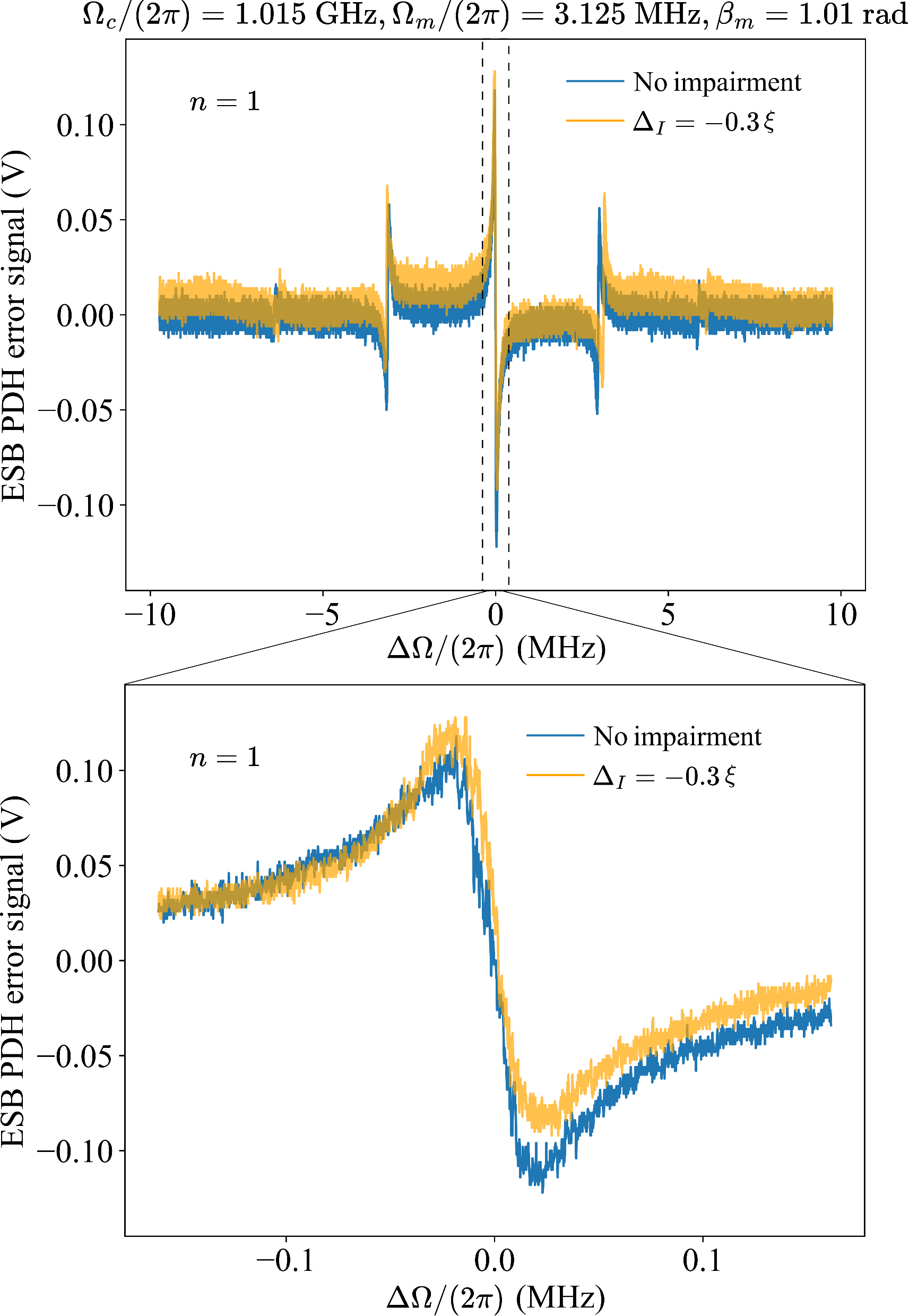}
    \caption{ESB error signals as a function of laser frequency detuning $\Delta\Omega$ measured with a $1112$~nm laser: the blue trace indicates the error signal generated with no I/Q impairment; the orange trace indicates the error signal generated with an injected impairment $\Delta_I=-0.3\xi$.}
    \label{fig:rfsoc_synth}
\end{figure}

We use the high-quality phase-modulated rf signals $V_{\mathrm{EOM}}(t)$ to generate ESB error signals using a PDH laser locking setup. We plot the resulting error signal as a function of the laser frequency detuning
\begin{equation}
\Delta\Omega \equiv \Omega_0+\Omega_c-2\pi N\times \mathrm{FSR}
\end{equation}
in Fig.~\ref{fig:rfsoc_synth}. The amplified phase-modulated rf signal from each design drives a fiber-coupled EOM (iXblue NIR-MPX-LN-05-00-P-P-FA-FA), which phase modulates laser light at $1112~\mathrm{nm}$ from a single-frequency distributed-feedback fiber laser (Koheras Adjustik Y10 system, NKT Photonics) with a free-running linewidth of $2\pi\times 2~\mathrm{kHz}$. As the frequency reference, we use a high-finesse ULE cavity (FSR $=1.5~\mathrm{GHz}$ and finesse $\simeq 6\times 10^{4}$) from Stable Laser Systems. We sweep the laser frequency to measure the ESB error signal.

In Fig.~\ref{fig:rfsoc_synth}, the ESB error signal is measured with $\beta_m=\beta_{\mathrm{opt}}$, $\Omega_m/(2\pi)=3.125~\mathrm{MHz}$, and $\Omega_c/(2\pi)=1.015~\mathrm{GHz}$. Uncompensated I/Q impairments produce a nonzero DC offset near resonance, i.e., a nonzero $\delta$ (see Sec.~\ref{sec:theory}). To illustrate this effect, we deliberately inject an impairment $\Delta_I=-0.3\xi$ into the rf signal and measure the ESB error signal. Fig.~\ref{fig:rfsoc_synth} compares the error signal with and without the injected $\Delta_I$: the impaired trace exhibits a clear offset near resonance, whereas no offset is observed for the compensated trace. More generally, we observe nonzero DC offsets when the I/Q impairments are not well compensated; their absence therefore indicates effective I/Q impairment compensation.

\begin{figure}[h]
    \centering
    \includegraphics[width=0.9\linewidth]{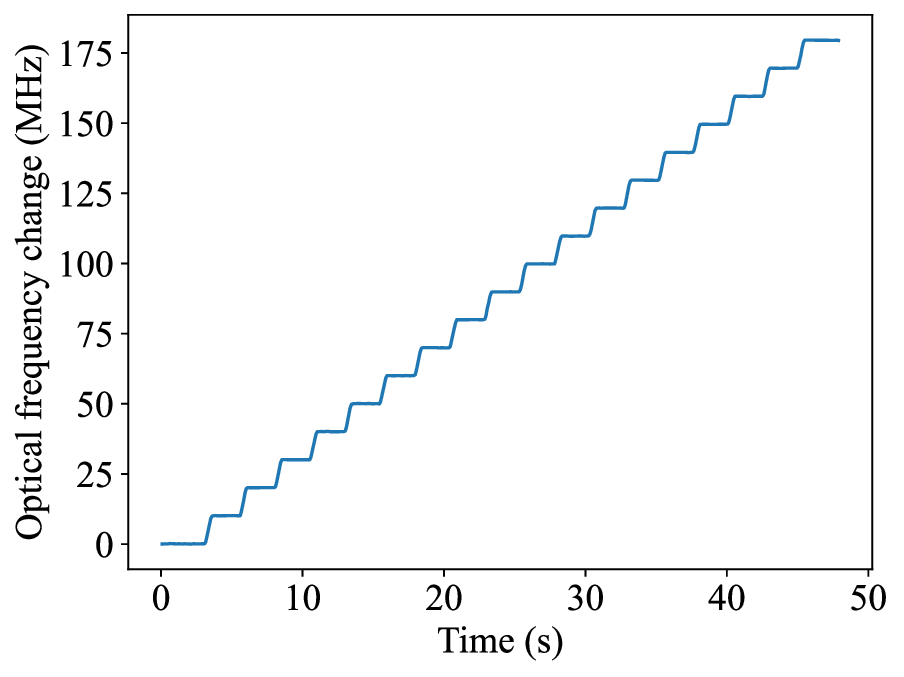}
    \caption{Frequency ramp of a $1112$~nm laser stabilized and tuned by the ESB PDH locking scheme. 
    }
    \label{fig:optical_freqramp}
\end{figure}

In addition, we demonstrate carrier-frequency tuning while maintaining the laser lock, using the optical setup described in Sec.~\ref{sec:optical}. We set the modulation frequency to $\Omega_m/(2\pi)=625~\mathrm{kHz}$ and the modulation depth to $\beta_m=1.01~\mathrm{rad}$. We initially set the carrier frequency to $\Omega_c/(2\pi)=807~\mathrm{MHz}$ and stabilize the laser. We then perform 18 frequency ramps of $10~\mathrm{MHz}$ with a step size of $10~\mathrm{Hz}$ and a $2~\mathrm{s}$ wait between consecutive ramps. The stabilized laser frequency is monitored with a wavemeter (HighFinesse Wavelength Meter WS Ultimate 2 MC). Fig.~\ref{fig:optical_freqramp} shows the measured frequency shifts during the ramps, confirming that the laser remains locked throughout the entire tuning sequence. Each ramp is completed in approximately $400~\mathrm{ms}$, consistent with the estimated FTW update rate of $380~\mathrm{ns}$. The laser maintains its lock throughout the frequency ramp, validating the continuous carrier-frequency tunability of our ESB locking device.

Although the frequency offset $\delta$ induced by injected I/Q impairments would ideally be quantified by directly measuring the optical frequency shift of the locked laser, this is challenging in our setup. The expected shifts (on the order of hundreds of Hz) are well below the typical $\sim 100~\mathrm{kHz}$ resolution of our wavemeter. While an optical beatnote measurement could resolve such shifts, it would require a second independent narrow-linewidth reference laser, which is not available in our laboratory. Consequently, measuring the offset in the PDH error signal provides our most direct and accessible method for characterizing shifts of the lock point.

\section{Conclusion and outlook}
In this work, we adapted QAM to synthesize the high-fidelity, phase-modulated rf drive required for ESB PDH laser locking. We developed a theoretical framework that connects intrinsic I/Q impairments to measurable I/Q errors and, critically, to distortions and lock-point offsets in the ESB error signal—providing practical bounds relevant to ultranarrow-linewidth stabilization. Guided by this model, we implemented a compact, direct-digital ESB signal generator on an UltraScale+ RFSoC platform and experimentally verified high-quality ESB waveforms with a large modulation index ($\beta_m=1.01$ rad) and low residual I/Q errors across a broad carrier-frequency range. Using these signals, we demonstrated ESB locking to a ULE reference cavity and phase-continuous, real-time carrier-frequency tuning that preserves lock during frequency ramps, enabling seamless laser tuning over a wide bandwidth. More broadly, the combination of a predictive impairment-to-offset model with a fully digital synthesis platform provides a practical route to robust long-term operation via calibration, predistortion, and closed-loop compensation of residual I/Q impairments, and offers a scalable approach for deploying widely tunable cavity-stabilized lasers in precision spectroscopy and metrology.

\begin{acknowledgments}
The authors thank Daniel Barker, Paul Lett, and Eric Benck for carefully reading the manuscript. 
This work was partially supported by the National Institute of Standards and Technology; the National Science Foundation through the Quantum Leap Challenge Institute for Robust Quantum Simulation (OMA-2120757); the Office of Naval Research (N000142212085); and the Air Force Office of Scientific Research Multidisciplinary University Research Initiative ``RAPSYDY in Q'' (FA9550-22-1-0339).
\end{acknowledgments}

\section*{Author Declarations}
\subsection*{Conflict of Interest}
The authors have no conflicts to disclose.


\section*{Data Availability}
The data that support the findings of this study are available from the corresponding author upon reasonable request.

\appendix
\section*{Appendix}
\section{Effect of I/Q impairments on the ESB error signal}
\label{sec:closedformequations}
\subsection{Analytical model}
In ESB locking, the ideal phase-modulated laser electric field $E^{\textrm{incident}}_{\textrm{ideal}}(t)$ incident on the reference cavity can be expressed as an infinite Fourier series using the Jacobi-Anger identity shown in Eq. \eqref{eq:idealESB}, where $J_n$ is the $\text{n}^\text{th}$ order Bessel function of the first kind. The amplitude of the laser sideband at frequency $\Omega_0 +n \Omega_c+k \Omega_m$ is $E_0J_n\left(\beta_c\right) J_k\left(n \beta_m\right)$. However, the inclusion of I/Q impairments significantly complicates the Fourier series expansion of the incident laser electric field $E^{\textrm{incident}}_{\textrm{impaired}}(q;t)$. Assuming that the amplitude of the impairment $(q-q_0)$ is small, we can Taylor expand $E^{\textrm{incident}}_{\textrm{impaired}}(q;t)$ in small $(q-q_0)$ as shown in Eqs. (\ref{eq:IQ error}) and (\ref{eq:impairedeqsupplementary}), where $\Lambda_{x,y}$ and $\Upsilon_{x,y}$ are the Fourier amplitudes of the corrections at $x\Omega_c+y\Omega_m$, and $q$ is the amplitude of the I/Q impairment under consideration: $\phi,g,\Delta_I$, and $\Delta_Q$. A second-order expansion is necessary for the closed-form expressions to accurately approximate the numerical results (see Figs.~\ref{fig:ESG} and~\ref{fig:ESO}).

\setcounter{equation}{0}
\begin{widetext}
\begin{minipage}{\textwidth}
\begin{align}
&E^{\textrm{incident}}_{\textrm{ideal}}(t)= E_0\rme^{j\Omega_0 t+j \beta_c \sin \left(\Omega_c t+\beta_m \sin \Omega_m t\right)} 
=  E_0\sum_n J_n\left(\beta_c\right) \rme^{j\Omega_0 t+j n \Omega_c t} \times \sum_k J_k\left(n \beta_m\right) \rme^{j k \Omega_m t} \nonumber\\
&=  E_0\sum_n \sum_k J_n\left(\beta_c\right) J_k\left(n \beta_m\right) \rme^{j\left (\Omega_0 +n \Omega_c+k \Omega_m\right) t},\label{eq:idealESB}\\
&E^{\textrm{incident}}_{\textrm{impaired}}(q;t)\simeq E^{\textrm{incident}}_{\textrm{impaired}}(t)\big|_{q=q_0}+(q-q_0)\frac{\partial E^{\textrm{incident}}_{\textrm{impaired}}(t)}{\partial q}\bigg|_{q=q_0}+\frac{(q-q_0)^2}{2!}\frac{\partial^2 E^{\textrm{incident}}_{\textrm{impaired}}(t)}{\partial q^2}\bigg|_{q=q_0}\nonumber\\
&=E^{\textrm{incident}}_{\textrm{ideal}}(t)\left(1\underbrace{-\frac{(q-q_0)^2}{2!}\beta_c^2 \left(\frac{\partial V_{\textrm{EOM}}(q;t)}{\partial q}\right)^2\bigg|_{q=q_0}}_{r(q;t)\cos(\eta(q;t))}+\right.\nonumber\\
&\qquad\left. j\underbrace{\beta_c\left((q-q_0) \frac{\partial V_{\textrm{EOM}}(q;t)}{\partial q}\bigg|_{q=q_0}+\frac{(q-q_0)^2}{2!} \frac{\partial^2 V_{\textrm{EOM}}(q;t)}{\partial q^2}\bigg|_{q=q_0}\right)}_{r(q;t)\sin(\eta (q;t))}\right)\label{eq:IQ error}\\
&=E_0\left(\sum_n \sum_k J_n\left(\beta_c\right) J_k\left(n \beta_m\right) \rme^{j\left(\Omega_0 +n \Omega_c+k \Omega_m\right) t}\right)\times\left(1+\sum_x \sum_y\left(j\beta_c\Lambda_{x,y}-\beta_c^2\Upsilon_{x,y}\right)\rme^{j\left(x \Omega_c+y \Omega_m\right)t}\right)\nonumber\\
&=E_0\bigg(\sum_n \sum_k J_n\left(\beta_c\right) J_k\left(n \beta_m\right) \rme^{j\left(\Omega_0 +n \Omega_c+k \Omega_m\right) t}\nonumber\\
&\qquad+\sum_l \sum_p\sum_x \sum_yJ_l\left(\beta_c\right) J_p\left(n \beta_m\right)\left(j\beta_c\Lambda_{x,y}-\beta_c^2\Upsilon_{x,y}\right)\rme^{j\left((x+l)\Omega_c+(y+p)\Omega_m\right)t}\bigg)\nonumber\\
&=E_0\sum_n \sum_k \rme^{j\left(\Omega_0 +n \Omega_c+k \Omega_m\right) t}\bigg(J_n\left(\beta_c\right) J_k\left(n \beta_m\right)+\sum_x \sum_yJ_{n-x}\left(\beta_c\right) J_{k-y}\left(n \beta_m\right)\left(j\beta_c\Lambda_{x,y}-\beta_c^2\Upsilon_{x,y}\right)\bigg)\label{eq:impairedeqsupplementary}
\end{align}
\begin{flalign}
E&^{\textrm{reflected}}_{\textrm{impaired}}(q;t)=E_0\sum_n \sum_k \rme^{j\left(\Omega_0 +n \Omega_c+k \Omega_m\right) t}F(\Omega_0 +n \Omega_c+k \Omega_m)\times\nonumber\\
&\qquad\bigg(J_n\left(\beta_c\right) J_k\left(n \beta_m\right)+\sum_x \sum_yJ_{n-x}\left(\beta_c\right) J_{k-y}\left((n-x) \beta_m\right)\left(j\beta_c\Lambda_{x,y}-\beta_c^2\Upsilon_{x,y}\right)\bigg)\label{eq:reflected}
\end{flalign}
\begin{equation}
\begin{aligned}
&\left|E^{\textrm{reflected}}_{\textrm{impaired}}(q;t)\right|^2\simeq |{E_0}|^2\sum_n \sum_k \sum_h \rme^{j(k-h) \Omega_m t}F(\Omega_0 +n\Omega_c+k \Omega_m)F^*(\Omega_0 +n\Omega_c+h \Omega_m)\times\\&\bigg(J_n\left(\beta_c\right) J_k\left(n \beta_m\right)+\sum_x \sum_yJ_{n-x}\left(\beta_c\right) J_{k-y}\left((n-x) \beta_m\right)\left(j\beta_c\Lambda_{x,y}-\beta_c^2\Upsilon_{x,y}\right)\bigg)\times\\&
\bigg(J_n\left(\beta_c\right) J_h\left(n \beta_m\right)-\sum_x \sum_yJ_{n-x}\left(\beta_c\right) J_{h-y}\left((n-x) \beta_m\right)\left(j\beta_c\Lambda^*_{x,y}+\beta_c^2\Upsilon^*_{x,y}\right)\bigg)\label{eq:measured}
\end{aligned}
\end{equation}
\end{minipage}
\end{widetext}
\setcounter{equation}{5}

We can cast Eq.~\eqref{eq:IQ error} in a more familiar form:

\begin{equation}
\begin{aligned}
E^{\textrm{incident}}_{\textrm{impaired}}(q;t)=E^{\textrm{incident}}_{\textrm{ideal}}(t)(1+r(q;t)\rme^{j\eta(q;t)}), 
\end{aligned}
\end{equation}
where
\begin{equation}
\begin{aligned}
r(q;t)\rme^{j\eta(q;t)}E^{\textrm{incident}}_{\textrm{ideal}}(t)
\end{aligned}
\end{equation}
is the instantaneous I/Q error vector in phasor form and $r(q;t)$ is the instantaneous error vector magnitude. When $|r(q;t)|\ll1$ and $|\eta(q;t)|\ll1$, the instantaneous I/Q magnitude error is $r(q;t)$ and the instantaneous I/Q phase error is $r(q;t)\eta(q;t)$. Therefore, reducing the I/Q magnitude error effectively minimizes the error vector magnitude.

Using Eq.~\eqref{eq:impairedeqsupplementary}, we can now derive the expression for the laser electric field reflected from the reference cavity $E^{\textrm{reflected}}_{\textrm{impaired}}(q;t)$ in Eq. \eqref{eq:reflected}, where $F(\Omega)$ is the cavity reflection coefficient~\cite{Black2001} for a laser sideband at frequency $\Omega$ and can be expressed as follows: 
\begin{align*}
    F(\Omega)=\frac{-2j(\Omega-2\pi N\times \textrm{FSR})/\kappa}{1+2j(\Omega-2\pi N\times \textrm{FSR})/\kappa}.
\end{align*}
 We assume that only the laser sideband at frequency $\Omega_0+\Omega_c$ is resonant with the reference cavity, in which case, 
\begin{align*}
    F(\Omega_0+n\Omega_c+h\Omega_m)&=\frac{-2j[(n-1)\Omega_c+h\Omega_m+\Delta\Omega]/\kappa}{1+2j[(n-1)\Omega_c+h\Omega_m+\Delta\Omega]/\kappa}\\
    &\approx-1+\delta_{n,1}\delta_{h,0}\left(1-\frac{2j\Delta\Omega/\kappa}{1+2j\Delta\Omega/\kappa}\right),
\end{align*}
 where $\Delta\Omega=\Omega_0+\Omega_c-2\pi N\times \textrm{FSR}$ for $\Omega_c,\Omega_m\gg\kappa$.

The signal measured at the photodetector is derived in Eq. \eqref{eq:measured}. For simplicity, we restrict ourselves to the $k\in\{-1,0,1\}$ and $h\in\{-1,0,1\}$ subspaces. The ESB error signal $V_{\textrm{error}}(q)$ is the amplitude of the sine quadrature of the photodetected signal:
\begin{equation}
    V_{\textrm{error}}(q)=2\int^{\pi/\Omega_m}_{-\pi/\Omega_m}\left|E^{\textrm{reflected}}_{\textrm{impaired}}(q;t)\right|^2 \sin(\Omega_m t) dt.
\end{equation}
Using $V_\textrm{error}(q)$, we can extract the closed-form expressions for $\gamma$ and $\delta$ as follows:
\begin{align}
    \text{$\gamma$}(q)&=\frac{\partial V_\textrm{error}(q)}{\partial\Delta\Omega}\bigg|_{\Delta\Omega=0}\\
    V_\textrm{error}(q)=0\implies \delta(q)&=\Delta \Omega(q).
\end{align}
We plot closed-form expressions as a function of the strength of the I/Q impairment amplitude $q$ in Fig.~\ref{fig:esberrors} in the main text.

\subsection{Numerical model}\label{sec:numericalmodel}
We numerically compute the $\gamma$ and $\delta$ under different types of I/Q impairments. First, we generate the modulated signal with impairments as an array of numbers. Then we demodulate the signal by integrating it with an array of numbers representing $\sin(\Omega_m t)$. The $\gamma$ and $\delta$ calculated from the demodulated error signal are plotted as points in Fig.~\ref{fig:esberrors}. The numerical calculation matches the analytical model well.

\section{On the carrier wave oscillator}
\label{sec:carrierwave}
As an alternative to the direct SDR approach presented in this work, ESB rf waveforms can be generated by I/Q-modulating a PLL-generated carrier using a baseband-sampling SDR (e.g., the ADALM-PLUTO platform briefly mentioned in Sec.~\ref{sec:design}). For ESB and DSB locking, the carrier-wave oscillator must simultaneously provide low phase noise and wideband frequency tunability, which are often conflicting requirements~\cite{Banerjee2017}. In practice, microwave carrier oscillators are therefore frequently stabilized with a PLL referenced to a low-drift source---typically an oven-controlled crystal oscillator (OCXO), an oven-controlled SAW oscillator (OCSO), or an atomic clock---to suppress long-term frequency drifts and improve phase-noise performance.

In baseband-sampling SDR modules such as ADALM-PLUTO, wide tuning ranges are typically achieved using integrated PLL/VCO chips that contain multiple silicon VCO cores with overlapping frequency subbands~\cite{Banerjee2017,Brennan2016,Banerjee2020,AdemAktas2004}. However, seamless phase-continuous tuning is generally only available within a single VCO subband (often on the order of tens of MHz, e.g., $\sim 40~\mathrm{MHz}$), since crossing between subbands requires switching VCO cores and retuning the PLL. As a result, such integrated PLL/VCO solutions are poorly matched to ESB locking, which benefits from few-GHz continuous carrier tuning without interruptions.

Discrete PLL+VCO architectures can provide seamless tuning over a few GHz~\cite{Harney2009,Collins2018,Brennan2016}, at the cost of increased system complexity and careful design of PLL dynamics (settling time) and closed-loop phase noise. Candidate oscillators include monolithic microwave integrated circuit (MMIC) VCOs, YIG (yttrium iron garnet) oscillators, dielectric resonator oscillators (DROs), and SAW oscillators, which can offer multi-GHz tuning ranges with favorable noise characteristics~\cite{Hartnagel2023,Fox2002}. In such designs, the oscillator choice is typically dictated by the required tuning range, the PLL settling time needed to maintain lock during tuning, and the achievable closed-loop phase-noise performance.

A common strategy to extend the effective tuning range of an integrated PLL/VCO is to tune within a single VCO subband and use frequency multiplication to reach the desired carrier band. This approach, however, can degrade phase noise because multiplication increases phase-noise power~\cite{Acar2021,Saavedra2011,Barrett1999}. An alternative is to generate the carrier directly with a numerically controlled oscillator (NCO) via direct digital synthesis (DDS), as in an RFSoC-based implementation. In DDS-based solutions, PLL settling-time and loop-filter stability concerns during tuning are eliminated, and sub-Hz frequency resolution and wideband, phase-continuous tuning are readily available~\cite{Acar2021,Cordesses2004,Devices1999,Calosso2012,Sander,Surber1996}.

 \begin{figure}[ht]
    \centering
    \includegraphics[width=\linewidth]{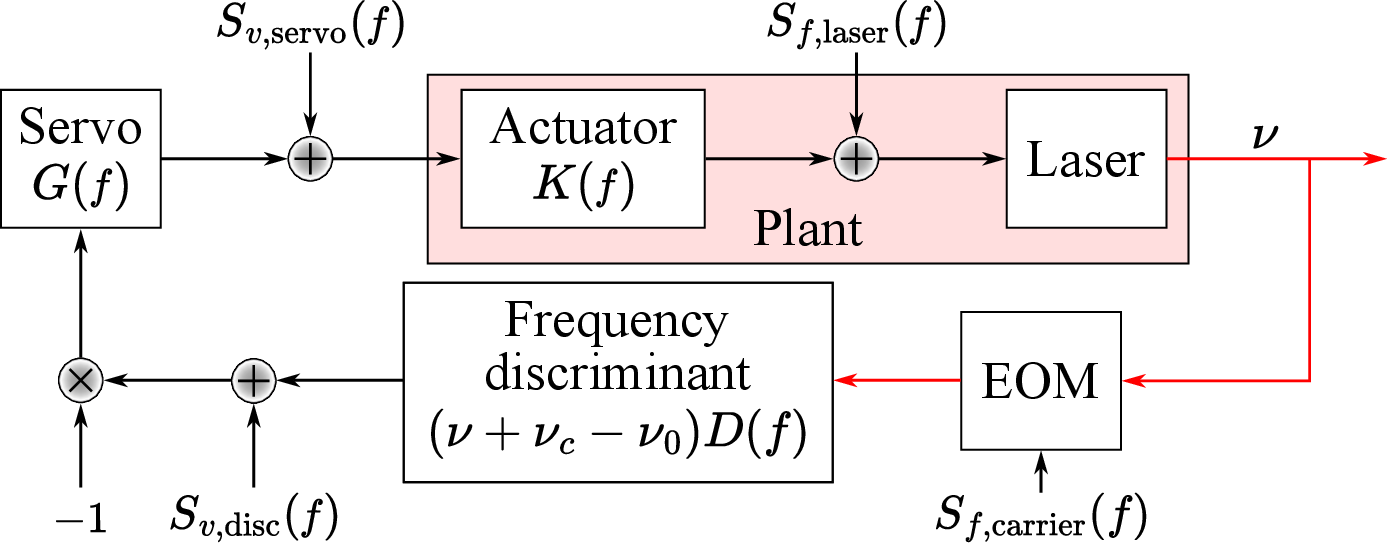}
    \caption{The feedback control system for offset sideband locking scheme in the presence of noise. The black arrows indicate the propagation of electronic signals, and the red arrows indicate the propagation of optical signals.}\label{fig:closedloopnoise}
\end{figure}

In offset-sideband locking, phase noise on the carrier-wave oscillator propagates directly to frequency noise of the stabilized laser. Fig.~\ref{fig:closedloopnoise} illustrates the feedback loop used for laser-frequency stabilization in this scheme~\cite{Subhankar_2024}. Here $G(f)$, $K(f)$, and $D(f)$ denote the transfer functions of the servo controller, the laser frequency actuator (e.g., cavity or current tuning), and the frequency discriminant (the PDH system), respectively. The terms $S(f)$ denote spectral densities of noise sources within the loop: the first subscript specifies the noise type, either voltage noise ($v$, in $\mathrm{V}/\sqrt{\mathrm{Hz}}$) or frequency noise ($f$, in $\mathrm{Hz}/\sqrt{\mathrm{Hz}}$), and the second subscript identifies the component injecting the noise.

The resulting closed-loop laser frequency-noise spectral density $S_{f,\mathrm{cl}}$ (in $\mathrm{Hz}/\sqrt{\mathrm{Hz}}$) is
{\footnotesize
\begin{equation}
S_{f,\mathrm{cl}}
=\frac{\sqrt{S_{f,\mathrm{laser}}^{2}
+\left|K S_{v,\mathrm{servo}}\right|^{2}
+\left|G K S_{v,\mathrm{disc}}\right|^{2}
+\left|G K D S_{f,\mathrm{carrier}}\right|^{2}}}
{\left|1+G K D\right|}.
\end{equation}
}%
In the large-loop-gain limit, $|G|\gg 1$, this expression reduces to
\begin{equation}\label{eq:noiselimit}
S_{f,\mathrm{cl}}
=\sqrt{\frac{S_{v,\mathrm{disc}}^{2}}{|D|^{2}}+S_{f,\mathrm{carrier}}^{2}}.
\end{equation}
Equation~\eqref{eq:noiselimit} makes explicit that carrier-oscillator phase noise (written here as an equivalent frequency-noise term $S_{f,\mathrm{carrier}}$) contributes directly to the residual frequency noise of the stabilized laser. Consequently, phase-noise contributions in the carrier-generation chain---including frequency multipliers---should be minimized to preserve laser-frequency stability.

\section*{References}
\bibliography{aipsamp}

@PREAMBLE{
 "\providecommand{\noopsort}[1]{}" 
 # "\providecommand{\singleletter}[1]{#1}%" 
}

@article{Katori2011,
author = {Katori, Hidetoshi},
doi = {10.1038/nphoton.2011.45},
issn = {17494885},
journal = {Nature Photonics},
number = {4},
pages = {203--210},
publisher = {Nature Publishing Group},
title = {Optical lattice clocks and quantum metrology},
volume = {5},
year = {2011}
}

@article{Derevianko2011,
archivePrefix = {arXiv},
arxivId = {1011.4622},
author = {Derevianko, Andrei and Katori, Hidetoshi},
doi = {10.1103/RevModPhys.83.331},
isbn = {0034-6861 1539-0756},
issn = {00346861},
journal = {Reviews of Modern Physics},
number = {2},
pages = {331--347},
title = {Colloquium: Physics of optical lattice clocks},
volume = {83},
year = {2011}
}

@article{Buikema2019,
author = {Aaron Buikema and Franklin Jose and Steven J. Augst and Peter Fritschel and Nergis Mavalvala},
journal = {Opt. Lett.},
number = {15},
pages = {3833--3836},
publisher = {Optica Publishing Group},
title = {Narrow-linewidth fiber amplifier for gravitational-wave detectors},
volume = {44},
month = {Aug},
year = {2019},
url = {https://opg.optica.org/ol/abstract.cfm?URI=ol-44-15-3833},
doi = {10.1364/OL.44.003833},
}

@article{Carlson2009,
author = {Carlson, Chad G. and Dragic, Peter D. and Price, R. Kirk and Coleman, J. J. and Swenson, Gary R.},
doi = {10.1109/JSTQE.2009.2012403},
issn = {1077260X},
journal = {IEEE Journal on Selected Topics in Quantum Electronics},
issue = {2},
pages = {451--461},
title = {A narrow-linewidth, {Yb} fiber-amplifier-based upper atmospheric {Doppler} temperature lidar},
volume = {15},
year = {2009}
}

@article{Tang2022,
author = {Liwei Tang and Liangbo Li and Jiachen Li and Minghua Chen},
journal = {Opt. Express},
number = {17},
pages = {30420--30429},
publisher = {Optica Publishing Group},
title = {Hybrid integrated ultralow-linewidth and fast-chirped laser for {FMCW LiDAR}},
volume = {30},
month = {Aug},
year = {2022},
url = {https://opg.optica.org/oe/abstract.cfm?URI=oe-30-17-30420},
doi = {10.1364/OE.465858},
}

@article{Bao2012,
author = {Bao, Xiaoyi and Chen, Liang},
doi = {10.3390/s120708601},
issn = {14248220},
journal = {Sensors (Switzerland)},
number= {7},
pages = {8601--8639},
title = {Recent Progress in Distributed Fiber Optic Sensors},
volume = {12},
year = {2012}
}

@article{Elsherif2022,
author = {Elsherif, Mohamed and Salih, Ahmed E. and Mu{\~{n}}oz, Monserrat Guti{\'{e}}rrez and Alam, Fahad and AlQattan, Bader and Antonysamy, Dennyson Savariraj and Zaki, Mohamed Fawzi and Yetisen, Ali K. and Park, Seongjun and Wilkinson, Timothy D. and Butt, Haider},
doi = {10.1002/adpr.202100371},
issn = {2699-9293},
journal = {Advanced Photonics Research},
number = {11},
title = {Optical Fiber Sensors: Working Principle, Applications, and Limitations},
volume = {3},
year = {2022},
pages = {2100371},
}

@Article{Sun2019,
AUTHOR = {Sun, Xiaochen and Zhang, Lingxuan and Zhang, Qihao and Zhang, Wenfu},
TITLE = {Si Photonics for Practical LiDAR Solutions},
JOURNAL = {Applied Sciences},
VOLUME = {9},
YEAR = {2019},
NUMBER = {20},
pages = {4225},
URL = {https://www.mdpi.com/2076-3417/9/20/4225},
ISSN = {2076-3417},
DOI = {10.3390/app9204225}
}

@article{Bai2021,
author = {Bai, Zhenxu and Zhao, Zhongan and Tian, Menghan and Jin, Duo and Pang, Yajun and Li, Sensen and Yan, Xiusheng and Wang, Yulei and Lu, Zhiwei},
title = {A comprehensive review on the development and applications of narrow-linewidth lasers},
journal = {Microwave and Optical Technology Letters},
volume = {64},
number = {12},
pages = {2244--2255},
doi = {https://doi.org/10.1002/mop.33046},
url = {https://onlinelibrary.wiley.com/doi/abs/10.1002/mop.33046},
year = {2022}
}

@article{Kapasi2020,
author = {D. P. Kapasi and J. Eichholz and T. McRae and R. L. Ward and B. J. J. Slagmolen and S. Legge and K. S. Hardman and P. A. Altin and D. E. McClelland},
journal = {Opt. Express},
number = {3},
pages = {3280--3288},
publisher = {Optica Publishing Group},
title = {Tunable narrow-linewidth laser at 2 {\textmu}m wavelength for gravitational wave detector research},
volume = {28},
month = {Feb},
year = {2020},
url = {https://opg.optica.org/oe/abstract.cfm?URI=oe-28-3-3280},
doi = {10.1364/OE.383685},
}

@article{Izumi2014,
author = {Kiwamu Izumi and Daniel Sigg and Lisa Barsotti},
journal = {Opt. Lett.},
number = {18},
pages = {5285--5288},
publisher = {Optica Publishing Group},
title = {Self-amplified lock of an ultra-narrow linewidth optical cavity},
volume = {39},
month = {Sep},
year = {2014},
url = {https://opg.optica.org/ol/abstract.cfm?URI=ol-39-18-5285},
doi = {10.1364/OL.39.005285},
}

@article{PhysRevA.107.L060801,
  title = {Observation of the $4{f}^{14}6{s}^{2} {}^{1}{S}_{0}\ensuremath{-}4{f}^{13}5d6{s}^{2}({J}=2)$ clock transition at 431 nm in ${}^{171}\mathrm{Yb}$},
  author = {Kawasaki, Akio and Kobayashi, Takumi and Nishiyama, Akiko and Tanabe, Takehiko and Yasuda, Masami},
  journal = {Phys. Rev. A},
  volume = {107},
  number = {6},
  pages = {L060801},
  numpages = {5},
  year = {2023},
  month = {Jun},
  publisher = {American Physical Society},
  doi = {10.1103/PhysRevA.107.L060801},
  url = {https://link.aps.org/doi/10.1103/PhysRevA.107.L060801}
}

@article{PhysRevLett.95.083003,
  title = {Observation and Absolute Frequency Measurements of the $^{1}{S}_{0}\mathrm{\text{\ensuremath{-}}}^{3}{P}_{0}$ Optical Clock Transition in Neutral Ytterbium},
  author = {Hoyt, C. W. and Barber, Z. W. and Oates, C. W. and Fortier, T. M. and Diddams, S. A. and Hollberg, L.},
  journal = {Phys. Rev. Lett.},
  volume = {95},
  number = {8},
  pages = {083003},
  numpages = {4},
  year = {2005},
  month = {Aug},
  publisher = {American Physical Society},
  doi = {10.1103/PhysRevLett.95.083003},
  url = {https://link.aps.org/doi/10.1103/PhysRevLett.95.083003}
}

@article{PhysRevA.98.022501,
  title = {Testing physics beyond the standard model through additional clock transitions in neutral ytterbium},
  author = {Dzuba, V. A. and Flambaum, V. V. and Schiller, S.},
  journal = {Phys. Rev. A},
  volume = {98},
  number = {2},
  pages = {022501},
  numpages = {12},
  year = {2018},
  month = {Aug},
  publisher = {American Physical Society},
  doi = {10.1103/PhysRevA.98.022501},
  url = {https://link.aps.org/doi/10.1103/PhysRevA.98.022501}
}

@article{PhysRevResearch.1.033113,
  title = {Isotope-shift spectroscopy of the ${}^{1}{S}_{0}\ensuremath{\rightarrow}{}^{3}{P}_{1}$ and ${}^{1}{S}_{0}\ensuremath{\rightarrow}{}^{3}{P}_{0}$ transitions in strontium},
  author = {Miyake, Hirokazu and Pisenti, Neal C. and Elgee, Peter K. and Sitaram, Ananya and Campbell, Gretchen K.},
  journal = {Phys. Rev. Res.},
  volume = {1},
  number = {3},
  pages = {033113},
  numpages = {8},
  year = {2019},
  month = {Nov},
  publisher = {American Physical Society},
  doi = {10.1103/PhysRevResearch.1.033113},
  url = {https://link.aps.org/doi/10.1103/PhysRevResearch.1.033113}
}

@article{Kohlhaas:12,
author = {Ralf Kohlhaas and Thomas Vanderbruggen and Simon Bernon and Andrea Bertoldi and Arnaud Landragin and Philippe Bouyer},
journal = {Opt. Lett.},
number = {6},
pages = {1005--1007},
publisher = {Optica Publishing Group},
title = {Robust laser frequency stabilization by serrodyne modulation},
volume = {37},
month = {Mar},
year = {2012},
url = {https://opg.optica.org/ol/abstract.cfm?URI=ol-37-6-1005},
doi = {10.1364/OL.37.001005},
}

@article{Zhou:23,
author = {Ocean Zhou and Andrew O. Neely and Zachary R. Pagel and Madeline Bernstein and Jack Roth and Holger Mueller},
journal = {Opt. Continuum},
number = {5},
pages = {1087--1092},
publisher = {Optica Publishing Group},
title = {Offset lock with a 440-{GHz} range using electro-optic modulation},
volume = {2},
month = {May},
year = {2023},
url = {https://opg.optica.org/optcon/abstract.cfm?URI=optcon-2-5-1087},
doi = {10.1364/OPTCON.474038},
}

@article{Ye:99,
author = {Jun Ye and John L. Hall},
journal = {Opt. Lett.},
number = {24},
pages = {1838--1840},
publisher = {Optica Publishing Group},
title = {Optical phase locking in the microradian domain: potential applications to {NASA} spaceborne optical measurements},
volume = {24},
month = {Dec},
year = {1999},
url = {https://opg.optica.org/ol/abstract.cfm?URI=ol-24-24-1838},
doi = {10.1364/OL.24.001838},
}

@book{Alencar2018,
author = {de Alencar, Marcelo Sampaio},
doi = {10.1201/9781003338864},
edition = {1st},
isbn = {9788770220262},
issn = {2196968X},
title = {Modulation Theory},
publisher ={River Publishers},
year = {2018}
}

@Book{Hanzo2011,
  author    = "Lajos L. Hanzo and Soon Xin Ng and Thomas Keller and William Webb",
  title     = {Quadrature Amplitude Modulation: From Basics to Adaptive Trellis-Coded, Turbo-Equalised and Space-Time Coded {OFDM, CDMA} and {MC-CDMA} Systems},
  publisher = "Wiley-IEEE Press",
  year      = "2004",
  doi = {10.1109/9780470010594},
  ISBN = {9780470094686}
}

@techreport{Nash2009,
  author      = "Nash, Eamon",
  title       = "Correcting Imperfections in {IQ} Modulators to Improve {RF} Signal Fidelity",
  institution = "Analog Devices",
  year        = "2009",
  type        = "Application Note",
  number      = "AN-1039",
  url = "https://www.analog.com/media/en/technical-documentation/application-notes/an-1039.pdf"
}

@article{Ghannouchi2013,
author = {Ghannouchi, Fadhel M. and Younes, Mayada and Rawat, Meenakshi},
doi = {10.1049/iet-map.2012.0663},
issn = {17518725},
journal = {IET Microwaves, Antennas and Propagation},
issu = {7},
pages = {518--534},
title = {Distortion and impairments mitigation and compensation of single- and multi-band wireless transmitters (invited)},
volume = {7},
year = {2013}
}

@ARTICLE{4389078,
  author={Angrisani, Leopoldo and D'Arco, Mauro and Vadursi, Michele},
  journal={IEEE Transactions on Instrumentation and Measurement}, 
  title={Clustering-Based Method for Detecting and Evaluating {I/Q} Impairments in Radio-Frequency Digital Transmitters}, 
  year={2007},
  volume={56},
  issu={6},
  pages={2139--2146},
  doi={10.1109/TIM.2007.908127}}

@ARTICLE{1275708,
  author={Georgiadis, A.},
  journal={IEEE Transactions on Vehicular Technology}, 
  title={Gain, phase imbalance, and phase noise effects on error vector magnitude}, 
  year={2004},
  volume={53},
  issu={2},
  pages={443--449},
  doi={10.1109/TVT.2004.823477}}

@article{Black2001,
author = {Black, Eric D.},
doi = {10.1119/1.1286663},
issn = {0002-9505, 1943-2909},
journal = {American Journal of Physics},
month = {jan},
issu = {1},
pages = {79--87},
title = {An introduction to {Pound–Drever–Hall} laser frequency stabilization},
url = {https://doi.org/10.1119/1.1286663},
volume = {69},
year = {2001}
}

@article{Mohammadian2021,
author = {Mohammadian, Amirhossein and Tellambura, Chintha},
doi = {10.1109/ACCESS.2021.3101845},
issn = {21693536},
journal = {IEEE Access},
pages = {111718--111791},
publisher = {IEEE},
title = {{RF} Impairments in Wireless Transceivers: Phase Noise, {CFO}, and {IQ} Imbalance - A Survey},
volume = {9},
year = {2021}
}

@article{Johnson:10,
author = {D. M. S. Johnson and J. M. Hogan and {S.-w.} Chiow and M. A. Kasevich},
journal = {Opt. Lett.},
issu = {5},
pages = {745--747},
publisher = {OSA},
title = {Broadband optical serrodyne frequency shifting},
volume = {35},
month = {Mar},
year = {2010},
url = {http://ol.osa.org/abstract.cfm?URI=ol-35-5-745},
doi = {10.1364/OL.35.000745},
}

@article{Ludlow2015,
  title = {Optical atomic clocks},
  author = {Ludlow, Andrew D. and Boyd, Martin M. and Ye, Jun and Peik, E. and Schmidt, P. O.},
  journal = {Rev. Mod. Phys.},
  volume = {87},
  number = {2},
  pages = {637--701},
  numpages = {65},
  year = {2015},
  month = {Jun},
  publisher = {American Physical Society},
  doi = {10.1103/RevModPhys.87.637},
  url = {https://link.aps.org/doi/10.1103/RevModPhys.87.637}
}

@article{Thorpe:08,
author = {J. I. Thorpe and K. Numata and J. Livas},
journal = {Opt. Express},
issu = {20},
pages = {15980--15990},
publisher = {OSA},
title = {Laser frequency stabilization and control through offset sideband locking to optical cavities},
volume = {16},
month = {Sep},
year = {2008},
url = {http://www.opticsexpress.org/abstract.cfm?URI=oe-16-20-15980},
doi = {10.1364/OE.16.015980},
}

@article{Houtz:09,
author = {Rachel Houtz and Cheong Chan and Holger M\"{u}ller},
journal = {Opt. Express},
issu = {21},
pages = {19235--19240},
publisher = {OSA},
title = {Wideband, Efficient Optical Serrodyne Frequency Shifting with a Phase Modulator and a Nonlinear Transmission Line},
volume = {17},
month = {Oct},
year = {2009},
url = {http://www.opticsexpress.org/abstract.cfm?URI=oe-17-21-19235},
doi = {10.1364/OE.17.019235},
}

@article{Drever1983,
  title = {Laser phase and frequency stabilization using an optical resonator},
  author = {Drever, R. W. P. and Hall, J. L. and Kowalski, F. V. and Hough, J. and Ford, G. M. and Munley, A. J. and Ward, H.},
  journal = {Applied Physics B},
  volume = {31},
  number = {2},
  pages = {97--105},
  numpages = {9},
  year = {1983},
  month = {Jun},
  publisher = {Springer},
  doi = {10.1007/BF00702605},
  url = {https://doi.org/10.1007/BF00702605},
}

@article{Bai_2017,
	doi = {10.1088/2040-8986/aa5a8c},
	url = {https://doi.org/10.1088/2040-8986/aa5a8c},
	year = 2017,
	month = {feb},
	publisher = {{IOP} Publishing},
	volume = {19},
	issu = {4},
	pages = {045501},
	author = {Jiandong Bai and Jieying Wang and Jun He and Junmin Wang},
	title = {Electronic sideband locking of a broadly tunable 318.6 nm ultraviolet laser to an ultra-stable optical cavity},
	journal = {Journal of Optics}
}

@article{Levine2018,
  title = {High-Fidelity Control and Entanglement of {Rydberg}-Atom Qubits},
  author = {Levine, Harry and Keesling, Alexander and Omran, Ahmed and Bernien, Hannes and Schwartz, Sylvain and Zibrov, Alexander S. and Endres, Manuel and Greiner, Markus and Vuleti\ifmmode \acute{c}\else \'{c}\fi{}, Vladan and Lukin, Mikhail D.},
  journal = {Phys. Rev. Lett.},
  volume = {121},
  number = {12},
  pages = {123603},
  numpages = {6},
  year = {2018},
  month = {Sep},
  publisher = {American Physical Society},
  doi = {10.1103/PhysRevLett.121.123603},
  url = {https://link.aps.org/doi/10.1103/PhysRevLett.121.123603}
}

@article{DeLeseleuc2018,
  title = {Analysis of imperfections in the coherent optical excitation of single atoms to {Rydberg} states},
  author = {de L\'es\'eleuc, Sylvain and Barredo, Daniel and Lienhard, Vincent and Browaeys, Antoine and Lahaye, Thierry},
  journal = {Phys. Rev. A},
  volume = {97},
  number = {5},
  pages = {053803},
  numpages = {9},
  year = {2018},
  month = {May},
  publisher = {American Physical Society},
  doi = {10.1103/PhysRevA.97.053803},
  url = {https://link.aps.org/doi/10.1103/PhysRevA.97.053803}
}

@article{Graham2019,
archivePrefix = {arXiv},
arxivId = {1908.06103},
author = {Graham, T. M. and Kwon, M. and Grinkemeyer, B. and Marra, Z. and Jiang, X. and Lichtman, M. T. and Sun, Y. and Ebert, M. and Saffman, M.},
doi = {10.1103/PhysRevLett.123.230501},
issn = {10797114},
journal = {Physical Review Letters},
issu = {23},
pages = {230501},
pmid = {31868460},
publisher = {American Physical Society},
title = {Rydberg-Mediated Entanglement in a Two-Dimensional Neutral Atom Qubit Array},
url = {https://doi.org/10.1103/PhysRevLett.123.230501},
volume = {123},
year = {2019}
}

@article{Alnis2008,
  title = {Subhertz linewidth diode lasers by stabilization to vibrationally and thermally compensated ultralow-expansion glass {Fabry-P}\'erot cavities},
  author = {Alnis, J. and Matveev, A. and Kolachevsky, N. and Udem, Th. and H\"ansch, T. W.},
  journal = {Phys. Rev. A},
  volume = {77},
  number = {5},
  pages = {053809},
  numpages = {9},
  year = {2008},
  month = {May},
  publisher = {American Physical Society},
  doi = {10.1103/PhysRevA.77.053809},
  url = {https://link.aps.org/doi/10.1103/PhysRevA.77.053809}
}

@article{Legaie:18,
author = {Remy Legaie and Craig J. Picken and Jonathan D. Pritchard},
journal = {J. Opt. Soc. Am. B},
issu = {4},
pages = {892--898},
publisher = {Optica Publishing Group},
title = {Sub-kilohertz excitation lasers for quantum information processing with {Rydberg} atoms},
volume = {35},
month = {Apr},
year = {2018},
url = {https://opg.optica.org/josab/abstract.cfm?URI=josab-35-4-892},
doi = {10.1364/JOSAB.35.000892},
}

@article{Rabga:23,
author = {T. Rabga and K. G. Bailey and M. Bishof and D. W. Booth and M. R. Dietrich and J. P. Greene and P. Mueller and T. P. O'Connor and J. T. Singh},
journal = {Opt. Express},
issu = {25},
pages = {41326--41338},
publisher = {Optica Publishing Group},
title = {Implementing an electronic sideband offset lock for isotope shift spectroscopy in radium},
volume = {31},
month = {Dec},
year = {2023},
url = {https://opg.optica.org/oe/abstract.cfm?URI=oe-31-25-41326},
doi = {10.1364/OE.500578},
}

@article{Gillot2022,
author = {Jonathan Gillot and Santerelli Falzon Tetsing-Talla and S\'{e}verine Denis and Gwenha\"{e}l Goavec-Merou and Jacques Millo and Cl\'{e}ment Lacro\^{u}te and Yann Kersal\'{e}},
journal = {Opt. Express},
number = {20},
pages = {35179--35188},
publisher = {Optica Publishing Group},
title = {Digital control of residual amplitude modulation at the {$10^{-7}$} level for ultra-stable lasers},
volume = {30},
month = {Sep},
year = {2022},
url = {https://opg.optica.org/oe/abstract.cfm?URI=oe-30-20-35179},
doi = {10.1364/OE.465597},
}

@article{Shi2018,
author = {Shi, Xiaohui and Zhang, Jie and Zeng, Xiaoyi and Lü, Xiaolong and Liu, Kui and Xi, Jing and Ye, Yanxia and Lu, Zehuang},
journal = {Applied Physics B},
pages = {153},
title = {Suppression of residual amplitude modulation effects in {Pound–Drever–Hall} locking},
volume = {124},
year = {2018},
url = {https://doi.org/10.1007/s00340-018-7021-y},
doi = {10.1007/s00340-018-7021-y}
}

@article{Milani:17,
author = {Gianmaria Milani and Benjamin Rauf and Piero Barbieri and Filippo Bregolin and Marco Pizzocaro and Pierre Thoumany and Filippo Levi and Davide Calonico},
journal = {Opt. Lett.},
issu = {10},
pages = {1970--1973},
publisher = {Optica Publishing Group},
title = {Multiple wavelength stabilization on a single optical cavity using  the offset sideband locking technique},
volume = {42},
month = {May},
year = {2017},
url = {https://opg.optica.org/ol/abstract.cfm?URI=ol-42-10-1970},
doi = {10.1364/OL.42.001970},
}

@article{Bridge:16,
author = {Elizabeth M. Bridge and Niamh C. Keegan and Alistair D. Bounds and Danielle Boddy and Daniel P. Sadler and Matthew P. A. Jones},
journal = {Opt. Express},
issu = {3},
pages = {2281--2292},
publisher = {Optica Publishing Group},
title = {Tunable cw {UV} laser with <35 {kHz} absolute frequency instability for precision spectroscopy of {Sr Rydberg} states},
volume = {24},
month = {Feb},
year = {2016},
url = {https://opg.optica.org/oe/abstract.cfm?URI=oe-24-3-2281},
doi = {10.1364/OE.24.002281},
}

@article{Guttridge2018,
  title = {Production of ultracold {C}${\mathrm{s}}^{*}\mathrm{Yb}$ molecules by photoassociation},
  author = {Guttridge, Alexander and Hopkins, Stephen A. and Frye, Matthew D. and McFerran, John J. and Hutson, Jeremy M. and Cornish, Simon L.},
  journal = {Phys. Rev. A},
  volume = {97},
  number = {6},
  pages = {063414},
  numpages = {8},
  year = {2018},
  month = {Jun},
  publisher = {American Physical Society},
  doi = {10.1103/PhysRevA.97.063414},
  url = {https://link.aps.org/doi/10.1103/PhysRevA.97.063414}
}

@article{Fox2002,
author = {Fox, Adrian},
journal = {Analog Dialogue},
volume = {36},
issue = {1},
pages = {13--16},
title = {{PLL} Synthesizers},
doi = "{ }",
url = {https://www.analog.com/media/en/analog-dialogue/volume-36/number-1/articles/volume36-number1.pdf},
year = {2002}
}

@article{Surber1996,
author = {Surber, Jim and McHugh, Leo},
journal = {Analog Dialogue},
issue = {3},
pages = {12--13},
title = {Single-chip direct digital synthesis vs. the analog {PLL}},
doi="{ }",
url = {https://www.analog.com/en/analog-dialogue/articles/dds-vs-analog-pll.html},
volume = {30},
year = {1996}
}

@techreport{Sander,
  author      = "Sander, Kay-Uwe",
  title       = "Frequency and Phase settling time measurements on {PLL} circuits",
  institution = "Rohde \& Schwarz",
  year        = "2018",
  type        = "Application Note",
  number      = "1EF102-1E",
  url = "https://scdn.rohde-schwarz.com/ur/pws/dl_downloads/dl_application/application_notes/1ef102/1EF102_1E_FrequencySettling.pdf"
}

@INPROCEEDINGS{Calosso2012,
  author={Calosso, Claudio E. and Gruson, Yannick and Rubiola, Enrico},
  booktitle={2012 IEEE International Frequency Control Symposium Proceedings}, 
  title={Phase noise and amplitude noise in {DDS}}, 
  year={2012},
  volume={},
  number={},
  pages={1-6},
  doi={10.1109/FCS.2012.6243619}}

@manual{Devices1999,
  title       = "A Technical Tutorial on Digital Signal Synthesis",
  organization = "Analog Devices",
  url = {http://www.ieee.li/pdf/essay/dds.pdf},
  year        = "1999"
}

@techreport{Technologies,
  author      = "{Keysight Technologies}",
  title       = "8 Hints for Making and Interpreting {EVM} Measurements",
  institution = "Keysight Technologies",
  year        = "2017",
  type        = "Application Note",
  number      = "5989-3144EN",
  url = "https://www.keysight.com/us/en/assets/7018-01305/application-notes/5989-3144.pdf"
}

@Book{Razavi2011,
  author    = "Behzad Razavi",
  title     = "RF Microelectronics",
  edition   =  "second",
  publisher = "Pearson",
  year      = "2011",
  ISBN      = "9780137134731",
  series    = "Prentice Hall Communications Engineering and Emerging Technologies Series"
}

@Book{Witte2001,
  author    = "Robert A. Witte",
  title     = "Spectrum and Network Measurements",
  publisher = "SciTech Publishing",
  year      = "2001",
  ISBN      = "9781884932168"
}

@Book{Banerjee2017,
  author    = "Banerjee, Dean",
  title     = "PLL Performance, Simulation, and Design",
  edition   =  "fifth",
  publisher = "Dog Ear Publishing",
  isbn      = "9781457551772",
  year      = "2017"
}

@techreport{Acar1,
  author      = "Acar, Erkan",
  title       = "How Error Vector Magnitude ({EVM}) Measurement Improves Your System-Level Performance",
  institution = "Analog Devices",
  year        = "2021",
  type        = "Technical Article",
  number      = "TA22797-7/21(A)",
  url = "https://www.analog.com/media/en/technical-documentation/tech-articles/how-evm-measurement-improves-system-level-performance.pdf"
}

@book{booksdr,
title = "Software Defined Radio with {Zynq Ultrascale+ RFSoC}",
editor = "Crockett, Louise H. and David Northcote and Robert W. Stewart",
year = "2023",
month = "jan",
day = "24",
isbn = "9781739588601",
url="https://www.rfsocbook.com/",
publisher = "Strathclyde Academic Media"
}

@book{AdemAktas2004,
author = {Aktas, Adem and Ismail, Mohammed},
doi = {10.1007/b117846},
publisher = {Springer New York, NY},
title = {{CMOS PLLs} and {VCOs} for {4G} Wireless},
year = {2004},
isbn={9781402080609}
}

@techreport{Banerjee2020,
  author      = "Banerjee, Dean and Mieso, Jacob",
  title       = "Dramatically Improve Your Lock Time with {VCO} Instant Calibration",
  institution = "Texas Instruments",
  year        = "2020",
  type        = "Application Report",
  number      = "SNAA342",
  url = "https://www.ti.com/lit/an/snaa342/snaa342.pdf"
}

@article{Collins2018,
  title = {Phase-Locked Loop {(PLL)} Fundamentals},
  author = {Collins, Ian},
  journal = {Analog Dialogue},
  volume = {52},
  issue = {3},
  pages = {13--18},
  year = {2018},
  publisher = {Analog Devices},
  doi="{ }",
  url = {https://www.analog.com/media/en/analog-dialogue/volume-52/number-3/volume52-number3.pdf}
}

@book{Hartnagel2023,
author = {Hartnagel, Hans L. and Quay, R{\"{u}}diger and Rohde, Ulrich L. and Rudolph, Matthias},
booktitle = {Fundamentals of RF and Microwave Techniques and Technologies},
doi = {10.1007/978-3-030-94100-0},
isbn = {9783030941000},
publisher = {Springer Cham},
title = {Fundamentals of {RF} and Microwave Techniques and Technologies},
year = {2023}
}

@article{Harney2009,
author = {Harney, Austin},
journal = {Analog Dialogue},
issue = {4},
pages = {13--16},
title = {Designing High-Performance Phase-Locked Loops with high voltage {VCOs}},
volume = {43},
year = {2011},
doi="{ }",
url={https://www.analog.com/media/en/analog-dialogue/volume-43/number-4/articles/volume43-number4.pdf}
}

@article{Cordesses2004,
author = {Cordesses, Lionel},
doi = {10.1109/MSP.2004.1311140},
issn = {10535888},
journal = {IEEE Signal Processing Magazine},
issue = {4},
pages = {50--54},
title = {Direct digital synthesis: A tool for periodic wave generation (Part 1 and 2)},
volume = {21},
year = {2004}
}

@techreport{Brennan2016,
  author      = "Brennan, Robert",
  title       = "Wideband Phase-locked Loops with Integrated Voltage Controlled Oscillators: Can They Replace a Discrete Solution?",
  institution = "Analog Devices",
  year        = "2016",
  type        = "Technical Article",
  number      = "TA14281-0-4/16",
  url = "https://www.analog.com/media/en/technical-documentation/tech-articles/wideband-phase-locked-loops-with-integrated-voltage-controlled-oscillators.pdf"
}

@techreport{Denisowski2022,
  author      = "Denisowski, Paul",
  title       = "Understanding {EVM}",
  institution = "Rohde \& Schwarz",
  year        = "2022",
  type        = "White Paper",
  number      = "PD 3683.8038.52",
  url = "https://cdn.rohde-schwarz.com.cn/pws/dl_downloads/premiumdownloads/premium_dl_pdm_downloads/3683_8038_52/Understanding-EVM_wp_en_3683-8038-52_v0100.pdf"
}

@techreport{NuWaves2019,
  author      = "{NuWaves Engineering}",
  title       = "Understanding Constellation Diagrams and How They Are Used",
  institution = "NuWaves Engineering",
  year        = "2019",
  type        = "Application Note",
  number      = "AN-005",
  url = "https://nuwaves.com/wp-content/uploads/AN-005-Constellation-Diagrams-and-How-They-Are-Used.pdf"
}

@manual{rfdc2024,
  title        = "{Zynq UltraScale+ RFSoC RF} Data Converter v2.6 {Gen 1/2/3/DFE LogiCORE IP} Product Guide",
  organization = "AMD",
  year         = "2024",
  url         = "https://docs.amd.com/r/en-US/pg269-rf-data-converter/Introduction"
}

@techreport{Acar2021,
  author      = "Acar, Erkan",
  title       = "Why a Fully Integrated Translation Loop Device Achieves the Best Phase Noise Performance",
  institution = "Analog Devices",
  year        = "2021",
  type        = "Technical Article",
  number      = "TA22936-4/21",
  url = "https://www.analog.com/media/en/technical-documentation/tech-articles/translation-loop-device-achieves-best-phase-noise-perform.pdf"
}

@incollection{Saavedra2011,
  author    = "Carlos E. Saavedra",
  title     = "Frequency Multiplier Design: Techniques and Applications",
  booktitle = "{CMOS} Nanoelectronics: Analog and {RF VLSI} Circuits",
  publisher = "McGraw-Hill Education",
  year      = "2011",
  editor    = "Iniewski, Krzysztof",
  pages     = "163--184",
  isbn     = "9780071755658",
  address   = "New York"
}

@techreport{Barrett1999,
  author      = " Curtis Barrett",
  title       = "Fractional/Integer-{N PLL} Basics",
  institution = "Texas Instruments",
  year        = "1999",
  type        = "Technical Brief",
  number      = "SWRA029",
  url = "https://www.ti.com/lit/an/swra029/swra029.pdf"
}

@article{Livas_2009,
doi = {10.1088/0264-9381/26/9/094016},
url = {https://dx.doi.org/10.1088/0264-9381/26/9/094016},
year = {2009},
month = {apr},
publisher = {},
volume = {26},
number = {9},
pages = {094016},
author = {Livas, J C and Thorpe, J I and Numata, K and Mitryk, S and Mueller, G and Wand, V},
title = {Frequency-tunable pre-stabilized lasers for {LISA} via sideband locking},
journal = {Classical and Quantum Gravity}
}

@article{PhysRevLett.121.053001,
  title = {Testing Quantum Electrodynamics in the Lowest Singlet State of Neutral Beryllium-9},
  author = {Cook, E. C. and Vira, A. D. and Patterson, C. and Livernois, E. and Williams, W. D.},
  journal = {Phys. Rev. Lett.},
  volume = {121},
  issue = {5},
  pages = {053001},
  numpages = {4},
  year = {2018},
  month = {Aug},
  publisher = {American Physical Society},
  doi = {10.1103/PhysRevLett.121.053001},
  url = {https://link.aps.org/doi/10.1103/PhysRevLett.121.053001}
}

@article{Sanjuan:21,
author = {Jose Sanjuan and Klaus Abich and Ludwig Bl\"{u}mel and Martin Gohlke and Vivek Gualani and Markus Oswald and Timm Wegehaupt and Thilo Schuldt and Claus Braxmaier},
journal = {Opt. Lett.},
number = {2},
pages = {360--363},
publisher = {Optica Publishing Group},
title = {Simultaneous laser frequency stabilization to an optical cavity and an iodine frequency reference},
volume = {46},
month = {Jan},
year = {2021},
url = {https://opg.optica.org/ol/abstract.cfm?URI=ol-46-2-360},
doi = {10.1364/OL.413419},
}

@article{10.1063/5.0076249,
    author = {Stefanazzi, Leandro and Treptow, Kenneth and Wilcer, Neal and Stoughton, Chris and Bradford, Collin and Uemura, Sho and Zorzetti, Silvia and Montella, Salvatore and Cancelo, Gustavo and Sussman, Sara and Houck, Andrew and Saxena, Shefali and Arnaldi, Horacio and Agrawal, Ankur and Zhang, Helin and Ding, Chunyang and Schuster, David I.},
    title = {The {QICK} (Quantum Instrumentation Control Kit): Readout and control for qubits and detectors},
    journal = {Review of Scientific Instruments},
    volume = {93},
    number = {4},
    pages = {044709},
    year = {2022},
    month = {04},
    issn = {0034-6748},
    doi = {10.1063/5.0076249},
    url = {https://doi.org/10.1063/5.0076249},
}

@article{Evered2025,
author = {Evered, Simon J. and Kalinowski, Marcin and Geim, Alexandra A. and Manovitz, Tom and Bluvstein, Dolev and Li, Sophie H. and Maskara, Nishad and Zhou, Hengyun and Ebadi, Sepehr and Xu, Muqing and Campo, Joseph and Cain, Madelyn and Ostermann, Stefan and Yelin, Susanne F. and Sachdev, Subir and Greiner, Markus and Vuletić, Vladan and Lukin, Mikhail D.},
journal = {Nature},
number = {8080},
pages = {341--347},
publisher = {Springer Nature Limited},
title = {Probing the {Kitaev} honeycomb model on a neutral-atom quantum computer},
volume = {645},
month = {Sep},
year = {2025},
url = {https://doi.org/10.1038/s41586-025-09475-0},
doi = {10.1038/s41586-025-09475-0},
}

@article{Carobene_2025,
doi = {10.1088/2058-9565/adcd97},
url = {https://dx.doi.org/10.1088/2058-9565/adcd97},
year = {2025},
month = {apr},
publisher = {IOP Publishing},
volume = {10},
number = {3},
pages = {035010},
author = {Carobene, Rodolfo and Candido, Alessandro and Serrano, Javier and Orgaz-Fuertes, Alvaro and Giachero, Andrea and Carrazza, Stefano},
title = {Qibosoq: an open-source framework for quantum circuit {RFSoC} programming},
journal = {Quantum Science and Technology}
}

@INPROCEEDINGS{10821231,
  author={Gartmann, Robert and Stumpert, Valentin and Scheller, Lukas and Weller, Richard and Ardila-Perez, Luis E. and Sander, Oliver},
  booktitle={2024 IEEE International Conference on Quantum Computing and Engineering (QCE)}, 
  title={Mixerless {RFSoC} Microwave Signal Generation for Superconducting Circuit Applications}, 
  year={2024},
  volume={02},
  number={},
  pages={565-566},
  doi={10.1109/QCE60285.2024.10407}}

@ARTICLE{10845073,
  author={Dudley, Tiamike and Plusquellic, Jim and Tsiropoulou, Eirini Eleni and Goldberg, Joshua and Stick, Daniel and Lobser, Daniel},
  journal={IEEE Transactions on Emerging Topics in Computing}, 
  title={Scatter-Gather {DMA} Performance Analysis Within an {SoC}-Based Control System for Trapped-Ion Quantum Computing}, 
  year={2025},
  volume={13},
  number={3},
  pages={841-852},
  doi={10.1109/TETC.2025.3528899}}

@article{PRXQuantum.5.020326,
  title = {Tunable Inductive Coupler for High-Fidelity Gates Between Fluxonium Qubits},
  author = {Zhang, Helin and Ding, Chunyang and Weiss, D.K. and Huang, Ziwen and Ma, Yuwei and Guinn, Charles and Sussman, Sara and Chitta, Sai Pavan and Chen, Danyang and Houck, Andrew A. and Koch, Jens and Schuster, David I.},
  journal = {PRX Quantum},
  volume = {5},
  issue = {2},
  pages = {020326},
  numpages = {18},
  year = {2024},
  month = {May},
  publisher = {American Physical Society},
  doi = {10.1103/PRXQuantum.5.020326},
  url = {https://link.aps.org/doi/10.1103/PRXQuantum.5.020326}
}

@article{PRXQuantum.5.030347,
  title = {Superconducting Qubits above 20 {GHz} Operating over 200 {mK}},
  author = {Anferov, Alexander and Harvey, Shannon P. and Wan, Fanghui and Simon, Jonathan and Schuster, David I.},
  journal = {PRX Quantum},
  volume = {5},
  issue = {3},
  pages = {030347},
  numpages = {19},
  year = {2024},
  month = {Sep},
  publisher = {American Physical Society},
  doi = {10.1103/PRXQuantum.5.030347},
  url = {https://link.aps.org/doi/10.1103/PRXQuantum.5.030347}
}

@article{Maetani_2024,
doi = {10.35848/1347-4065/ad40ea},
url = {https://dx.doi.org/10.35848/1347-4065/ad40ea},
year = {2024},
month = {jul},
publisher = {IOP Publishing},
volume = {63},
number = {7},
pages = {078001},
author = {Maetani, Kazunori and Machino, Akinori and Koike, Keisuke and Morisaka, Shinichi and Miyanishi, Koichiro and Kobayashi, Toshiki and Toyoda, Kenji and Negoro, Makoto and Miyoshi, Takefumi and Ohira, Ryutaro},
title = {Application of {RFSoC}-based arbitrary waveform generator for coherent control of atomic qubits},
journal = {Japanese Journal of Applied Physics}
}

@article{Zeng:21,
author = {Yong Zeng and Zhuo Fu and Yang-Yang Liu and Xiao-Dong He and Min Liu and Peng Xu and Xiao-Hong Sun and Jin Wang},
journal = {Appl. Opt.},
number = {5},
pages = {1159--1163},
publisher = {Optica Publishing Group},
title = {Stabilizing a laser frequency by the {Pound-Drever-Hall} technique with an acousto-optic modulator},
volume = {60},
month = {Feb},
year = {2021},
url = {https://opg.optica.org/ao/abstract.cfm?URI=ao-60-5-1159},
doi = {10.1364/AO.415011},
}

@article{Baryshev_2012,
doi = {10.1070/QE2012v042n04ABEH014646},
url = {https://doi.org/10.1070/QE2012v042n04ABEH014646},
year = {2012},
month = {apr},
publisher = {},
volume = {42},
number = {4},
pages = {315},
author = {Baryshev, Vyacheslav N},
title = {Laser frequency stabilisation by the {Pound—Drever-Hall} method using an acousto-optic phase modulator operating in the pure {Raman—Nath} diffraction regime},
journal = {Quantum Electronics}
}

@article{Hildebrand:25,
author = {Roame A. Hildebrand and Wance Wang and Connor Goham and Alessandro Restelli and Joseph W. Britton},
journal = {Opt. Express},
number = {25},
pages = {51842--51851},
publisher = {Optica Publishing Group},
title = {Spectrally-pure optical serrodyne modulation for continuously-tunable laser offset locking},
volume = {33},
month = {Dec},
year = {2025},
url = {https://opg.optica.org/oe/abstract.cfm?URI=oe-33-25-51842},
doi = {10.1364/OE.569114},
}

@article{Hildebrand:25-1,
author = {Roame A. Hildebrand and Wance Wang and Connor Goham and Alessandro Restelli and Joseph W. Britton},
journal = {Opt. Express},
number = {22},
pages = {45886--45893},
publisher = {Optica Publishing Group},
title = {Errors in {PDH} offset locking due to spurious spectral features},
volume = {33},
month = {Nov},
year = {2025},
url = {https://opg.optica.org/oe/abstract.cfm?URI=oe-33-22-45886},
doi = {10.1364/OE.572595},
}

@article{Takamoto2003,
  title = {Spectroscopy of the ${^{1}S_{0}\mathrm{\text{\ensuremath{-}}}^{3}P_{0}}$ Clock Transition of $^{87}\mathrm{S}\mathrm{r}$ in an Optical Lattice},
  author = {Takamoto, Masao and Katori, Hidetoshi},
  journal = {Phys. Rev. Lett.},
  volume = {91},
  issue = {22},
  pages = {223001},
  numpages = {4},
  year = {2003},
  month = {Nov},
  publisher = {American Physical Society},
  doi = {10.1103/PhysRevLett.91.223001},
  url = {https://link.aps.org/doi/10.1103/PhysRevLett.91.223001}
}

@article{Nicholson2015,
  title = {Systematic evaluation of an atomic clock at $2 \times 10^{-−18}$ total uncertainty},
  author = {Nicholson, T.L. and Campbell, S.L. and Hutson, R.B. and Marti, G.E. and Bloom, B.J. and McNally, R.L. and Zhang, W. and Barrett, M.D. and Safronova, M.S. and Strouse, G.F. and Tew, W.L. and Ye, J.},
  journal = {Nature Communications},
  volume = {6},
  issue = {1},
  pages = {6896},
  numpages = {4},
  year = {2015},
  month = apr,
  doi = {10.1038/ncomms7896},
  url = {https://doi.org/10.1038/ncomms7896}
}

@article{Dolde2025,
  title = {Direct measurement of the ${^{3}P_{0}}$ clock state natural lifetime in $^{87}\mathrm{Sr}$},
  author = {Dolde, Jonathan and Ganapathy, Dhruva and Zheng, Xin and Ma, Shuo and Beloy, Kyle and Kolkowitz, Shimon},
  journal = {Phys. Rev. A},
  volume = {112},
  issue = {2},
  pages = {023121},
  numpages = {8},
  year = {2025},
  month = {Aug},
  publisher = {American Physical Society},
  doi = {10.1103/f6pt-flnt},
  url = {https://link.aps.org/doi/10.1103/f6pt-flnt}
}

@article{Muniz2021,
  title = {{Cavity-QED} measurements of the $^{87}\mathrm{Sr}$ millihertz optical clock transition and determination of its natural linewidth},
  author = {Muniz, Juan A. and Young, Dylan J. and Cline, Julia R. K. and Thompson, James K.},
  journal = {Phys. Rev. Res.},
  volume = {3},
  issue = {2},
  pages = {023152},
  numpages = {21},
  year = {2021},
  month = {May},
  publisher = {American Physical Society},
  doi = {10.1103/PhysRevResearch.3.023152},
  url = {https://link.aps.org/doi/10.1103/PhysRevResearch.3.023152}
}

@article{Porsev2004,
  title = {Possibility of an optical clock using the $6{}^{1}{S}_{0}\ensuremath{\rightarrow}6{}^{3}{P}_{0}^{o}$ transition in ${}^{171,173}\mathrm{Yb}$ atoms held in an optical lattice},
  author = {Porsev, Sergey G. and Derevianko, Andrei and Fortson, E. N.},
  journal = {Phys. Rev. A},
  volume = {69},
  issue = {2},
  pages = {021403},
  numpages = {4},
  year = {2004},
  month = {Feb},
  publisher = {American Physical Society},
  doi = {10.1103/PhysRevA.69.021403},
  url = {https://link.aps.org/doi/10.1103/PhysRevA.69.021403}
}

@article{Rosenband2007,
  title = {Observation of the {$^{1}S_{0}\ensuremath{\rightarrow}{}^{3}P_{0}$} Clock Transition in $^{27}\mathrm{Al}^{+}$},
  author = {Rosenband, T. and Schmidt, P. O. and Hume, D. B. and Itano, W. M. and Fortier, T. M. and Stalnaker, J. E. and Kim, K. and Diddams, S. A. and Koelemeij, J. C. J. and Bergquist, J. C. and Wineland, D. J.},
  journal = {Phys. Rev. Lett.},
  volume = {98},
  issue = {22},
  pages = {220801},
  numpages = {4},
  year = {2007},
  month = {May},
  publisher = {American Physical Society},
  doi = {10.1103/PhysRevLett.98.220801},
  url = {https://link.aps.org/doi/10.1103/PhysRevLett.98.220801}
}

@phdthesis{Subhankar_2024,
  title={Engineering optical lattices for ultracold atoms with spatial features and periodicity below the diffraction limit and dual-species optical tweezer arrays for rubidium and ytterbium for {R}ydberg-interaction-mediated quantum simulations},
  url={https://drum.lib.umd.edu/handle/1903/33341},
  doi={10.13016/1Q83-PKJW},
  school={University of Maryland},
  author={Subhankar, Sarthak},
  year={2024},
  address={College Park, MD},
}

@article{Wang2025,
  title = {A practical guide to feedback control for {Pound–Drever–Hall} laser linewidth narrowing},
  author = {Wang, Wance and Subhankar, Sarthak and Britton, Joseph W.},
  journal = {Applied Physics B},
  volume = {131},
  issue = {7},
  pages = {146},
  year = {2025},
  month = jun,
  publisher = {American Physical Society},
  doi = {10.1007/s00340-025-08490-3},
  url = {https://doi.org/10.1007/s00340-025-08490-3}
}

@ARTICLE{Armstrong1936,
  author={Armstrong, E.H.},
  journal={Proceedings of the Institute of Radio Engineers}, 
  title={A Method of Reducing Disturbances in Radio Signaling by a System of Frequency Modulation}, 
  year={1936},
  volume={24},
  number={5},
  pages={689-740},
  doi={10.1109/JRPROC.1936.227383}}

\end{document}